\documentclass[usegraphicx]{mn2e}
\title[\xmm{} and the Pleiades]
{\xmm{} and the Pleiades -- I: Bright coronal sources and the X-ray
emission from intermediate-type stars}

\author[Briggs \& Pye]{K.R.\ Briggs$^{1,2}$\thanks{briggs@astro.phys.ethz.ch} and J.P.\ Pye$^1$\thanks{pye@star.le.ac.uk}\\
$^1$X-Ray Astronomy Group, Department of Physics and Astronomy, 
	University of Leicester, Leicester LE1 7RH, UK\\
$^2$Paul Scherrer Institut, CH-5232 Villigen PSI, Switzerland\\
}
\date{ Accepted 2003 July 2nd }

\newcommand{\et}{et~al.\ }
\newcommand{\lx}{\mbox{$L_{\rm X}$}}
\newcommand{\lbol}{\mbox{$L_{\rm bol}$}}

\newcommand{\kms}{\mbox{${\rm km}\, {\rm s}^{-1}$}}
\newcommand{\ergs}{\mbox{${\rm erg}\,{\rm s}^{-1}$}}
\newcommand{\chisq}{\mbox{$\chi^2$}}
\newcommand{\lxlbol}{$L_{\rm X}/L_{\rm bol}$}
\newcommand{\loglxlbol}{$\log (L_{\rm X}/L_{\rm bol})$}
\newcommand{\loglx}{$\log L_{\rm X}$}
\newcommand{\nh}{$N_{\rm H}$}
\newcommand{\vsini}{$v \sin i$}
\newcommand{\mekal}{{\sc mekal}}
\newcommand{\zsol}{$Z_{\odot}$}
\newcommand{\zfree}{$Z_{\rm free}$}
\newcommand{\onesig}{1-$\sigma$}
\newcommand{\quies}{quasi-steady}
\newcommand{\Quies}{Quasi-steady}

\newcommand{\ei}{{\it Einstein}}
\newcommand{\ro}{{\it ROSAT}}
\newcommand{\asca}{{\it ASCA}}

\newcommand{\chandra}{{\it Chandra}}
\newcommand{\xmm}{{\it XMM-Newton}}
\newcommand{\pn}{pn}

\newcommand{\aap}{A\&A}
\newcommand{\aaps}{A\&AS}
\newcommand{\aj}{AJ}

\newcommand{\apj}{ApJ}
\newcommand{\apjs}{ApJS}
\newcommand{\araa}{ARAA}
\newcommand{\mnras}{MNRAS}

\newcommand{\pasp}{PASP}
\begin{document}

\maketitle
\begin{abstract}
We present results of X-ray spectral and timing analyses of solar-like 
(spectral types F5--K8) and intermediate-type (B4--F4) Pleiads observed in a
$40$-ks \xmm{} EPIC exposure, probing X-ray luminosities (\lx) 
up to a factor 10 fainter than previous studies using the
\ro{} PSPC. All 8 solar-like members have ``\quies'' \lx$\:\ga 10^{29}$
\ergs{} consistent with the known rotation--activity relation, and 4
exhibit flares. Using a hydrodynamic modelling technique we derive loop
half-lengths ${\cal L}\, \la 0.5\, R_{\star}$ 
for the two strongest flares, on HII~1032 and HII~1100. 
Near the beginning of its flare, HII~1100's lightcurve shows a
feature with a profile suggestive of a total occultation of the flaring
loop. Eclipse by a substellar companion in a close orbit is possible
but would seem an extraordinarily fortuitous event; absorption by a
fast-moving cloud of cool material requires \nh{} at least two orders of
magnitude greater than any solar or stellar
prominence. An occultation may have been mimicked by the
coincidence of two flares, though the first, its decay time
shorter than its rise time and suggestive of ${\cal L} \sim 0.02\,
R_{\star}$, would be unusual. 

Spectral modelling of the \quies{}
emission shows a rising trend in coronal temperature from F and
slowly-rotating G stars through K stars to fast-rotating G stars, and
a preference for low coronal metallicity. These features are consistent with
those of nearby solar-like stars, although none of the three stars
showing ``saturated'' emission bears the significant component at 2
keV seen in the saturated coronae of AB~Dor and 47~Cas.
Of 5 intermediate-type stars, 2 are undetected (\lx$\:< 4 \times
10^{27}$ \ergs) and 3 show X-ray emission with a spectrum and \lx{}
consistent with origin from an active solar-like companion.
\end{abstract}

\begin{keywords}
X-rays: stars -- stars: activity -- stars: coronae -- stars:
early-type -- stars: late-type
-- open clusters and associations: individual: the Pleiades 
\end{keywords}

\section{Introduction}
\label{intro}

The Pleiades is the archetypal stellar sample with age $\approx 100$
Myr and near-solar photospheric metallicity. Relative youth,
proximity, compactness and richness make it key to understanding stellar
X-ray emission and its evolution. Cluster members cover spectral types
from late-B to late-M, enabling the simultaneous study of the
different processes driving X-ray emission from stars of different
internal structure. 

\begin{figure*}
\centering{
\hbox{
\includegraphics[height=5.9cm, origin=c, angle=0]{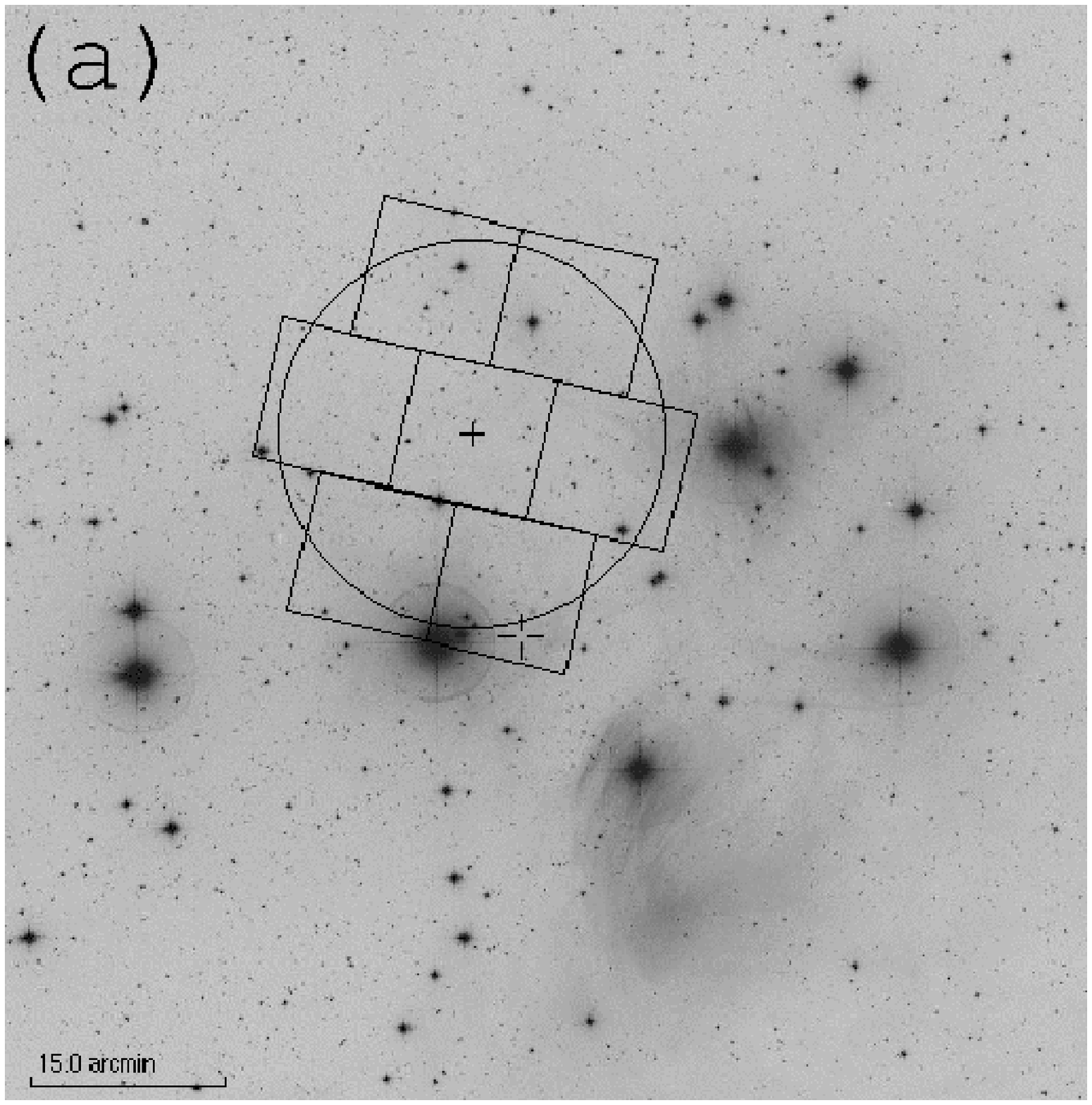}
\hspace{-0.2cm}
\includegraphics[height=5.9cm, origin=c, angle=0]{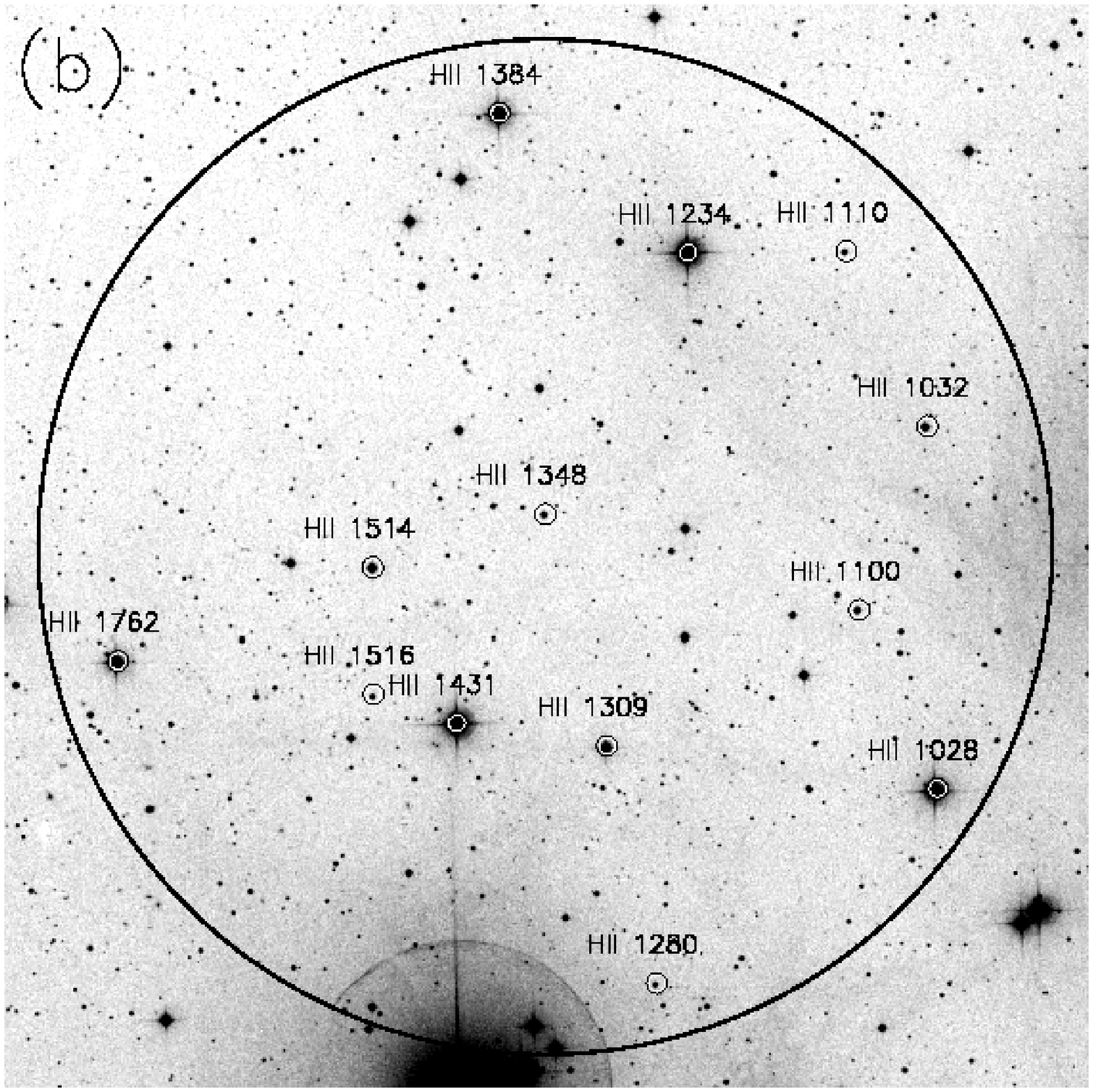}
\hspace{-0.2cm}
\includegraphics[height=5.9cm, origin=c, angle=0]{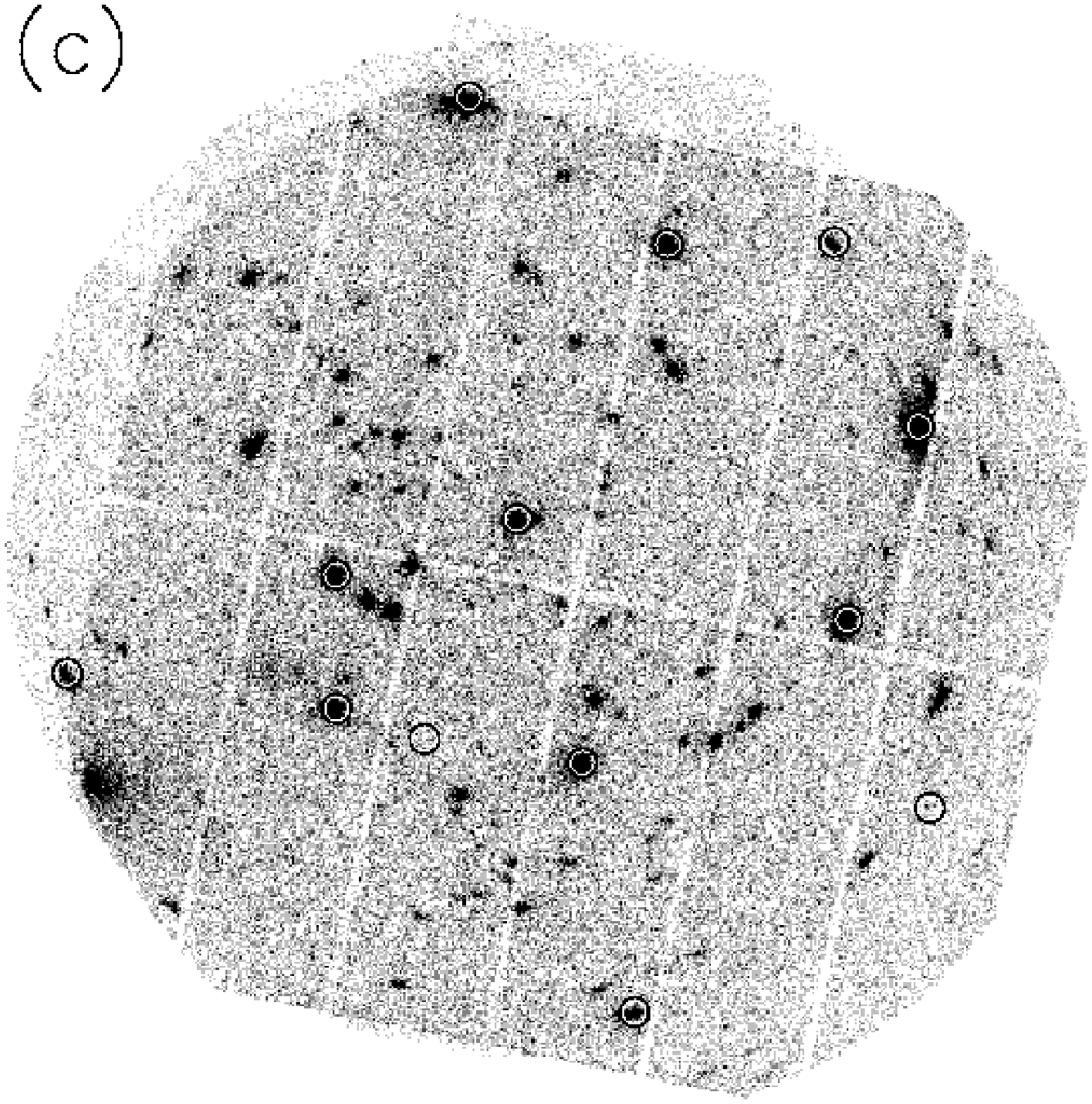}
}
\caption{Images of the Pleiades \xmm{} field. (a) Position of the
EPIC FOV in the Pleiades (Digitized Sky Survey (DSS)
image; overlaid MOS2 field from ALADIN). (b) DSS optical
image. The large circle marks EPIC's 30 arcmin diameter FOV. (c)
Mosaiced EPIC (MOS1+MOS2+\pn) X-ray (0.3--4.5 keV) 
image. In (b) and (c), the positions of intermediate-type and
solar-like Pleiads are circled.   
}
\label{fig_imgs}
}
\end{figure*}

Solar-like stars (spectral types $\approx$ F5--M3) have
radiative cores and convective outer envelopes, and X-ray emission is
believed to be from hot plasma confined in 
magnetic loops in the corona, strong magnetic field being generated by
a dynamo located at the base of the convection zone
($\alpha\Omega$ dynamo: Parker 1975), as on the Sun. 
Supportive correlations of increasing coronal X-ray and chromospheric line
(e.g. H$\alpha$, Ca II H, K) luminosities with increasing convection
zone depth and increasing rotation rates have been well-observed
(Pallavicini \et 1981; Noyes 
\et 1984; Pizzolato \et 2003), although the saturation of these
``magnetic activity indicators'' at high rotation rates, at an
X-ray-to-bolometric luminosity ratio of \lxlbol{} $\approx 10^{-3}$
(Vilhu 1984), awaits a full explanation. X-ray luminosities, \lx, 
may exceed $10^{3}$ times that of the Sun at the peak of its activity
cycle and younger stellar samples show higher mean
\lx{} as rotation rates slow with time through the braking effect of
a magnetized stellar wind. The youth of the Pleiades ensures a number
of ``ultra-fast-rotating'' solar-like members (UFRs), showing saturated
levels of magnetic activity, of which there are few examples in the
nearby field.

The Sun's X-ray-emitting corona is highly variable and
highly-structured. Stellar X-ray emission is also highly variable,
exhibiting flares with profiles echoing those seen on
the Sun, though reaching peak \lx{} up to $10^{3}$ times higher. X-ray
telescopes cannot resolve stellar coronae into active regions and
individual loops as seen on the Sun, but loop sizes may be inferred
from analysis of flare decays (Reale \& Micela 1998). Loop lengths a
large fraction of, but smaller than, the stellar
radius have been derived from flares observed on a small number of
nearby solar-like stars (e.g. Maggio \et 2000; G\"udel \et
2001; Covino \et 2001). Younger, faster-rotating, samples show
flares of higher peak \lx{} and with higher frequency (Stelzer, 
Neuh\"auser \& Hambaryan 2000), hence the Pleiades is a good target
for the study of stellar flares and coronal structure.

The temperature structure and composition of some stellar coronae 
diverge from those of the solar corona.
Therein, the bulk of X-ray-emitting plasma is at 1--2 MK
(e.g. Orlando, Peres \& Reale 2000), though flaring plasma is often as
hot as 10 MK (Reale, Peres \& Orlando 2001), while 
faster-rotating stars have more plasma at higher temperatures (e.g. G\"udel,
Guinan \& Skinner 1997), with the most active stars having much plasma
at 20 MK (e.g. Mewe \et 1996).
Elemental abundances in the solar corona differ from those in its
photosphere (so-called \emph{solar abundances}), elements with low
first ionization potential (FIP), such as Fe, appearing \emph{overabundant}
%\emph{the FIP effect}: 
(Meyer 1985), although variability occurs with
location within the corona (McKenzie \& Feldman 1992) and during
flares (Reames, Meyer \& von Rosenvinge 1994). In the coronae of
highly-active solar-like stars, however, low-FIP elements appear
\emph{underabundant} (e.g. Singh \et 1999), and the FIP-dependence of
coronal abundances changes according to activity level
(Audard \& G\"udel 2002). The proximity of the Pleiades
enables the study of temperature structure and composition in coronae
of a number of individual solar-like stars, helping to extend the
relatively small sample of coronae in which these physical conditions have
been measured.

While solar-like stars have hot coronae and early-type ($\la$ B3) stars are
thought to generate X-rays via hot shocks within their massive
stellar winds, stars of intermediate type
($\approx$B4--F4) lack both a deep
convective envelope and a massive stellar wind and are thus generally
considered incapable of strong X-ray emission: any observed is 
conventionally attributed to a later-type companion with a hot corona
(e.g. Golub \et 1983; Micela \et 1996; 1999), although in most cases
such a companion remains undiscovered. Intermediate-type Pleiads,
relatively close and well-studied for binarity, provide a good sample
in which to test this assertion. 
 
The X-ray emission from low-mass stars (spectral types $\ga$ M3),
which are thought to be fully-convective and hence unable to
support an $\alpha\Omega$ dynamo, is examined in a companion paper
(Briggs \& Pye 2003).

The core of the Pleiades has been extensively surveyed at X-ray
wavelengths by the \ei{} (Caillault \& Helfand 1985; Micela \et 1990) 
and \ro{} (Stauffer \et 1994; Micela \et 1996; Stelzer \& Neuh\"auser
2001; Micela \et 1999) observatories, and a single field therein has
been observed twice by \chandra{} (Krishnamurthi \et 2001; Daniel,
Linsky \& Gagn\'e 2002; henceforth DLG02).

\begin{table*}
\begin{minipage}{\textwidth}
\caption{
Data for Pleiades members in the EPIC
field. Columns show: 
%(1) HII number; 
(2) spectral type (SIMBAD), 
(3) and (4) RA and Dec (J2000) from USNO-A2.0 (Monet \et 1998); 
(5) and (6) $V$ (Belikov \et 1998) and 2MASS $J$ magnitudes (Cutri
\et 2000); 
(7) $\log$ \lbol{} in \ergs;
(8) stellar radius in $10^{10}$ cm;
(9) proper motion membership probability (Belikov \et 1998); 
(10) and (11) projected rotational velocity in km s$^{-1}$ and reference: 
1. Soderblom \et 1993; 2. Queloz \et 1998; 3. Terndrup \et 2000; 
(12) and (13) rotation period in d and reference:
4. Marilli, Catalano \& Frasca 1997; 5. Krishnamurthi \et 1998; ``:'' indicates $P_{\rm rot}/\sin i$ has been calculated from \vsini{} and $R_{\star}$;
(14) flag noting spectroscopic (SB), visual (VB) and suspected
photometric (Ph?) binaries (references in \S~\protect\ref{sec_res}, Table~\protect\ref{tbl_early}).
}
%\vspace{0.1cm}
\label{tbl_opt}
\footnotesize
\centering{
\begin{tabular}{llllrrrrcrclcl}
\hline
\multicolumn{1}{c}{HII} & 
\multicolumn{1}{c}{SpT} & 
\multicolumn{1}{c}{RA} & 
\multicolumn{1}{c}{Dec} & 
\multicolumn{1}{c}{$V$} & 
\multicolumn{1}{c}{$J$} & 
\multicolumn{1}{c}{[$L_{\rm bol}$]} &
\multicolumn{1}{c}{$R_{\star}$} &
\multicolumn{1}{c}{$P_{\rm mem}$} & 
\multicolumn{1}{c}{$v \sin i$} & 
\multicolumn{1}{c}{Ref.} & 
\multicolumn{1}{c}{$P_{\rm rot}$} & 
\multicolumn{1}{c}{Ref.} & 
\multicolumn{1}{c}{Bin.}\\
\multicolumn{1}{c}{(1)} & 
\multicolumn{1}{c}{(2)} & 
\multicolumn{1}{c}{(3)} &  
\multicolumn{1}{c}{(4)} &  
\multicolumn{1}{c}{(5)} & 
\multicolumn{1}{c}{(6)} & 
\multicolumn{1}{c}{(7)} & 
\multicolumn{1}{c}{(8)} & 
\multicolumn{1}{c}{(9)} & 
\multicolumn{1}{c}{(10)} & 
\multicolumn{1}{c}{(11)} & 
\multicolumn{1}{c}{(12)} & 
\multicolumn{1}{c}{(13)} & 
\multicolumn{1}{c}{(14)}\\
\hline
1234 & B9.5V& 3 46 59.40 & +24 31 12.4 &  6.82 &  6.67 & 35.15 & 13.3 & 0.23 & 260 & 1 & 0.37 & : & VB \\
1431 & A0V  & 3 47 29.45 & +24 17 18.0 &  6.81 &  6.61 & 35.15 & 13.3 & 0.61 &  40 & 1 & 2.4  & : & SB \\
1028 & A2V  & 3 46 27.10 & +24 15 21.2 &  7.35 &  7.11 & 34.85 & 11.8 & 0.73 & 110 & 1 & 0.78 & : & VB \\
1384 & A4V  & 3 47 23.97 & +24 35 20.0 &  7.66 &  7.12 & 34.70 & 11.3 & 0.83 & 215 & 1 & 0.38 & : & Ph?\\
1762 & A9V  & 3 48 13.50 & +24 19 07.5 &  8.27 &  7.50 & 34.43 & 10.6 & 0.87 & 180 & 1 & 0.43 & : & SB, VB\\
1309 & F6V  & 3 47 09.98 & +24 16 37.9 &  9.46 &  8.53 & 33.97 &  8.7 & 0.70 &  85 & 1 & 0.74 & : &    \\
1514 & G5   & 3 47 40.38 & +24 21 54.6 & 10.48 &  9.32 & 33.59 &  6.8 & 0.65 &  14 & 2 & 3.33 & 4 &    \\
1032 & G8V  & 3 46 28.36 & +24 26 04.2 & 11.34 &  9.62 & 33.28 &  5.8 & 0.95 &  37 & 2 & 1.31 & 5 & Ph?\\
1100 & K3V  & 3 46 37.22 & +24 20 38.8 & 12.16 & 10.12 & 33.03 &  5.1 & 0.35 &   5 & 2 & 7.4  & : & VB \\
1348 & K5   & 3 47 17.99 & +24 23 28.9 & 12.61 & 10.38 & 32.89 &  4.8 & 0.01 &   5 & 2 & 6.9  & : & SB \\
1110 & K6.5e& 3 46 38.83 & +24 31 15.1 & 13.29 & 11.04 & 32.69 &  4.2 & 0.98 &   6 & 2 & 5.1  & : &    \\
1516 & K    & 3 47 40.30 & +24 18 09.3 & 14.02 & 11.14 & 32.47 &  3.7 & 0.75 & 105 & 2 & 0.25 & : & Ph?\\
1280 & K8   & 3 47 03.52 & +24 09 37.0 & 14.55 & 11.63 & 32.31 &  3.3 & 0.64 &  85 & 3 & 0.30 & 5 &    \\
%\\
\hline
\end{tabular}
}
\vspace{-0.25cm}
\end{minipage}
\end{table*}

The \ro{} surveys detected X-ray emission, attributed to the coronae of late-type companions, from $\approx$ a third of
intermediate-type stars (Stauffer \et 1994; Micela \et 1996).
Krishnamurthi \et (2001) judged the soft, bright X-ray emission
of four intermediate-type stars in the \chandra{} field to
be intrinsic to these stars, but their known binarity led DLG02 to
implicate solar-like secondaries.

The \ro{} surveys indicated X-rays were emitted by practically all 
solar-like Pleiads. Gagn\'e, Caillault \& Stauffer (1995; henceforth
GCS95) analysed the 
individual spectra of X-ray-bright Pleiads (\loglx $\ga 30.0$) and the
composite spectra of samples subdivided by spectral type (F, G, K, M) and
into slow (\vsini{} $< 16$ \kms) and fast (\vsini{}
$> 16$ \kms) rotators. Only among G stars did rotation rate seem to effect
significant spectral differences, but there was a rising sequence in
temperature from F through slowly-rotating G then K and M
to fast-rotating G. GCS95 did not investigate 
coronal abundances due to the low energy-resolution of the PSPC, but
DLG02 found evidence for low Fe-abundance in the
coronae of three K-stars. 

Flares identified in the \ro{} surveys showed \lx{} increases of factors
2--40, peak \loglx{} of 29.8--31.2, and decay time-scales of 1--13 ks,
and were more frequent on K- and M-stars than on G-stars, which have
lower mean \lxlbol{} 
(GCS95; Stelzer \et 2000). GCS95 inferred a loop half-length of ${\cal L} \ga
6\, R_{\star}$ (stellar radii), for a large flare on the K-star
HII~1516 using a quasi-static cooling loop model, but this model is
not self-consistent in its assumptions and overestimates the loop
length (Reale 2002), the energy range of the PSPC was insufficient to
constrain the peak temperature, and the lightcurve was sparsely sampled.

In comparison to \ro{}, the greater sensitivity, wider energy range,
better energy resolution and continuous time coverage of \xmm{}'s EPIC 
detectors enable spectral studies at higher resolution of Pleiades
members with lower X-ray luminosities, the detection of smaller
flares, and better-constrained modelling of large flares. 
We use the \xmm{} EPIC cameras to perform spectral and timing analyses
of stars of spectral types B--K in a single 15 arcmin-radius field in
the core of the Pleiades, with the aims:
(a) to estimate temperature structure and metallicity and, through
analysis of the decays of large flares, loop size in
the coronae of individual solar-like Pleiads and hence compare these
properties to those of well-studied coronae of nearby solar-like
stars; and (b) to see if X-ray emission from intermediate-type stars
is consistent with emission from the coronae of later-type
companions. 
DLG02 addressed similar questions using their 60-ks \chandra{} study
of a neighbouring field. While the better angular 
resolution of \chandra{} offered lower background count-rates, our
\xmm{} observation adds to this study by sampling a more-complete
range of spectral types, providing 
more counts per source to enable better-constrained spectral
modelling, expanding the sample of well-studied X-ray-detected
intermediate-type Pleiads and being able to analyse large flares.

The paper is organized as 
follows: \S~2 describes the membership, optical and physical
data on the Pleiades used in this work; \S~3 details the X-ray
observations and data analysis performed; \S~4 reports and discusses
our results; \S~5 summarizes the work.

\section{Pleiades membership and physical parameters}
\label{sec_memb}

\begin{figure}
\centering{
\includegraphics[width=7cm, angle=270]{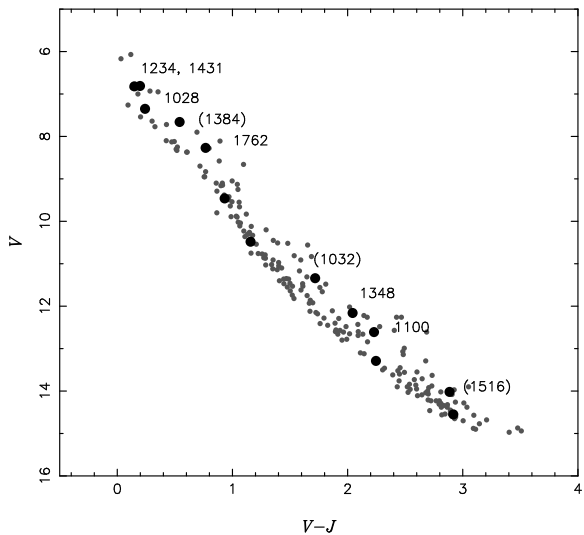}
\caption{Colour-magnitude diagram for Pleiades members -- in (black) and
outside (grey) the EPIC FOV -- indicating
confirmed (bracketless) and suspected photometric (bracketed)
binaries. 
%$V$ is from Belikov \et (1998); $J$ is from the
%2MASS Second Incremental Data Release (Cutri \et 2000).
}
\label{fig_cmd}
}
\end{figure}

We have drawn a list of solar-like and intermediate-type Pleiades
members from the
catalogue of Belikov \et (1998), whose survey used optical ($BVR$)
photometry, complete to $V \approx 17$, and
proper-motion measurements, with a baseline of 23 or 30 y, to
determine membership. Henceforth we use the terms ``Pleiads'' or
``members'' to refer to stars included in this catalogue.
The \xmm{} EPIC field of view (FOV), its position within the Pleiades
shown in Fig.~\ref{fig_imgs}a, contains 13 such members,
whose individual positions in an optical image
of the field are marked in Fig.~\ref{fig_imgs}b. There are
examples of each spectral type within the range of interest: 1 B-,
4 A-, 1 F-, 2 G-, and 5 K-stars. These members are identified throughout
the paper using HII numbers (table~2 of Hertzsprung 1947). Positional,
photometric, rotational, binarity and membership data (where
available) have been culled from the literature and are listed in
Table~\ref{tbl_opt}. The sample includes one G-type and two K-type
UFRs. We have further used a colour-magnitude diagram
(Fig.~\ref{fig_cmd}) to highlight suspected
photometric binaries. We adopt a distance to the Pleiades of 127 pc, 
$(m-M)_{0} = 5.52$, (e.g. Stello \& Nissen \et 2001) as
used in previous X-ray surveys. Extinction, $A_{V} = 0.12$,
and reddening, $E(B-V) = 0.04$, to the cluster are
fairly uniform across the EPIC field (Stauffer 1984), and imply a
hydrogen column density of \nh{} $= 2 \times 10^{20}$ cm$^{-2}$ using
the relation of Paresce (1984). The metallicity 
of the cluster is near-solar (e.g. King \et 2000). We have inferred bolometric
luminosities and stellar radii (the latter estimated through the
Stefan-Boltzmann 
law) by interpolation, using $M_{V}$, of the Girardi \et (2000) models
for stars with solar metallicity and age 100 Myr.

\section{X-ray observations and data analysis}
\label{sec_xdata}

Observation 0094780101 was a Guaranteed Time pointing (PI: M.~Watson)
%(for the \xmm{} Survey Science Centre) 
centred on the brown dwarf Teide 1 (J2000: $\alpha=03$~47~18.0,
$\delta=+24$~22~31), performed on 1st Sep 2000, in orbit 134. An
exposure time of 50 ks was scheduled but only 40.6 ks was achieved in
the EPIC \pn{} (Str{\" u}der \et 2001) and 33.0 ks in each of the
two EPIC MOS (Turner \et 2001) detectors. The Thick filter was used
in each of the three telescopes.
%Further details of the observation are summarised in Table~\ref{tbl_obs}.

The data were processed using the Science Analysis System (SAS)
v5.2\footnote{http://xmm.vilspa.esa.es/.}. We filtered each eventlist 
to exclude `bad' events (using the \#XMMEA\_EM and
\#XMMEA\_EP filters for MOS and \pn{} respectively), uncalibrated event
patterns (PATTERN $> 12$ for MOS; PATTERN $> 4$ for \pn), and
visually-identified ``hot'' pixels. Several
short background flares occurred but had no significant effect on the
analysis of bright sources so affected periods were only excluded in
the estimation of upper limits for undetected sources.

Fig.~\ref{fig_imgs}c shows a mosaic of images from all three EPIC
cameras in the energy range 0.3--4.5 keV. All 8 solar-like stars
(F--K), and 3 of the 5 intermediate-type stars (B--A) were clear 
by-eye detections. We henceforth refer to these 11 stars collectively as 
``X-ray-bright Pleiads''. The SAS source-detection software was run on images 
from the individual cameras, finding the best-fitting X-ray position
to be within 3.5 arcsec of the optical position (USNO-A2.0: Monet \et
1998) for each X-ray-bright Pleiad, and finding the remaining two
intermediate-type Pleiads, HII~1028 and 1431, to be 
undetected at a maximum likelihood threshold of $M\!L > 12$. 
Upper limits (at the 95 per
cent confidence level) to their X-ray luminosities were
estimated from the counts in the 0.3--4.5 keV band within 16 arcsec of
the optical position (using the ``classical'' prescription of 
Kraft, Burrows \& Nousek (1991) and the algebraic approximations to the
Poissonian upper limit derived by Gehrels (1986)) by correcting for
enclosed energy and effective exposure time, and assuming a 0.8 keV
\mekal{} source spectrum and distance of 127 pc. 

We performed spectral and timing analyses of the X-ray-bright
Pleiads using primarily data from the \pn, as its higher sensitivity
and longer exposure time collected typically a factor 3--4 more counts
per source than a single MOS. MOS data were used where they provided
$\ga 500$ source counts and were not affected by CCD edges (HII~1384, 1762). 
For each X-ray-bright Pleiad, we extracted a spectrum and lightcurve
from a circular region centred on the best-fitting X-ray source 
position. The radius of this circle, $r$, ranged from 30 to 60
arcsec and was constrained by two concerns: (a) contamination by
neighbouring sources; (b) retention of a good signal-to-noise 
ratio. The former provided the stronger constraint in most cases. 
In each case, a background spectrum and lightcurve were
also extracted from a surrounding annulus, with inner radius $\ge 1.5\,
r$ and area $\ga 5$ times the source 
extraction area after areas contaminated by other 
sources had been removed. The number of
background counts, $B$, expected to fall within each source extraction region 
was estimated from background maps produced by the source
detection prodecure and each background product was scaled to
contribute $B$ counts to its associated source
product. Analysis of the source products is described below.

\subsection{Time series analysis}
\label{sec_xdata_lc}

\begin{figure*}
\centering{
\begin{minipage}[b]{.45\textwidth}
  \centering
  \includegraphics[height=7.1cm,angle=270]{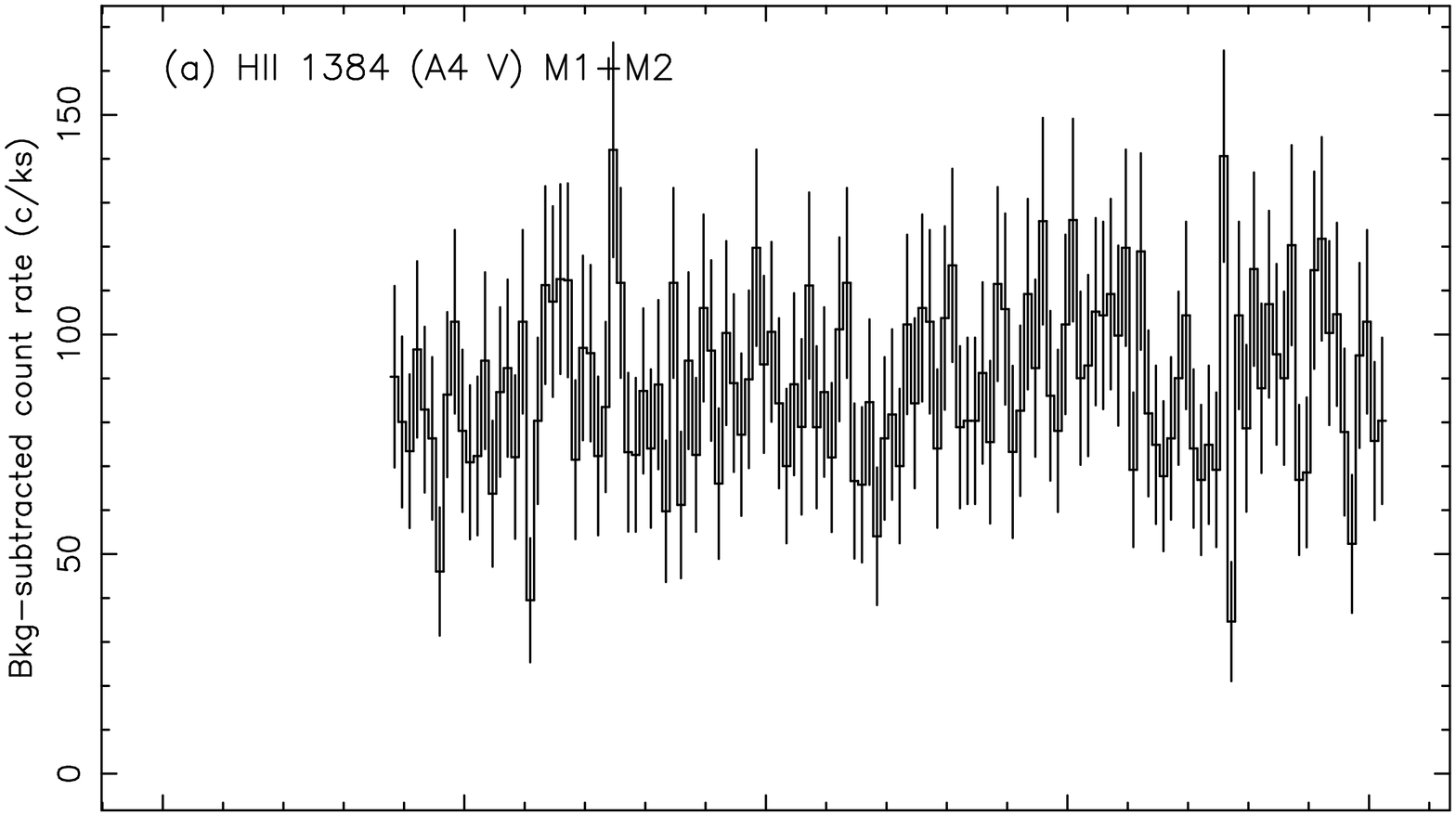}
\end{minipage}%
\begin{minipage}[b]{.45\textwidth}
  \centering
  \includegraphics[height=7.1cm,angle=270]{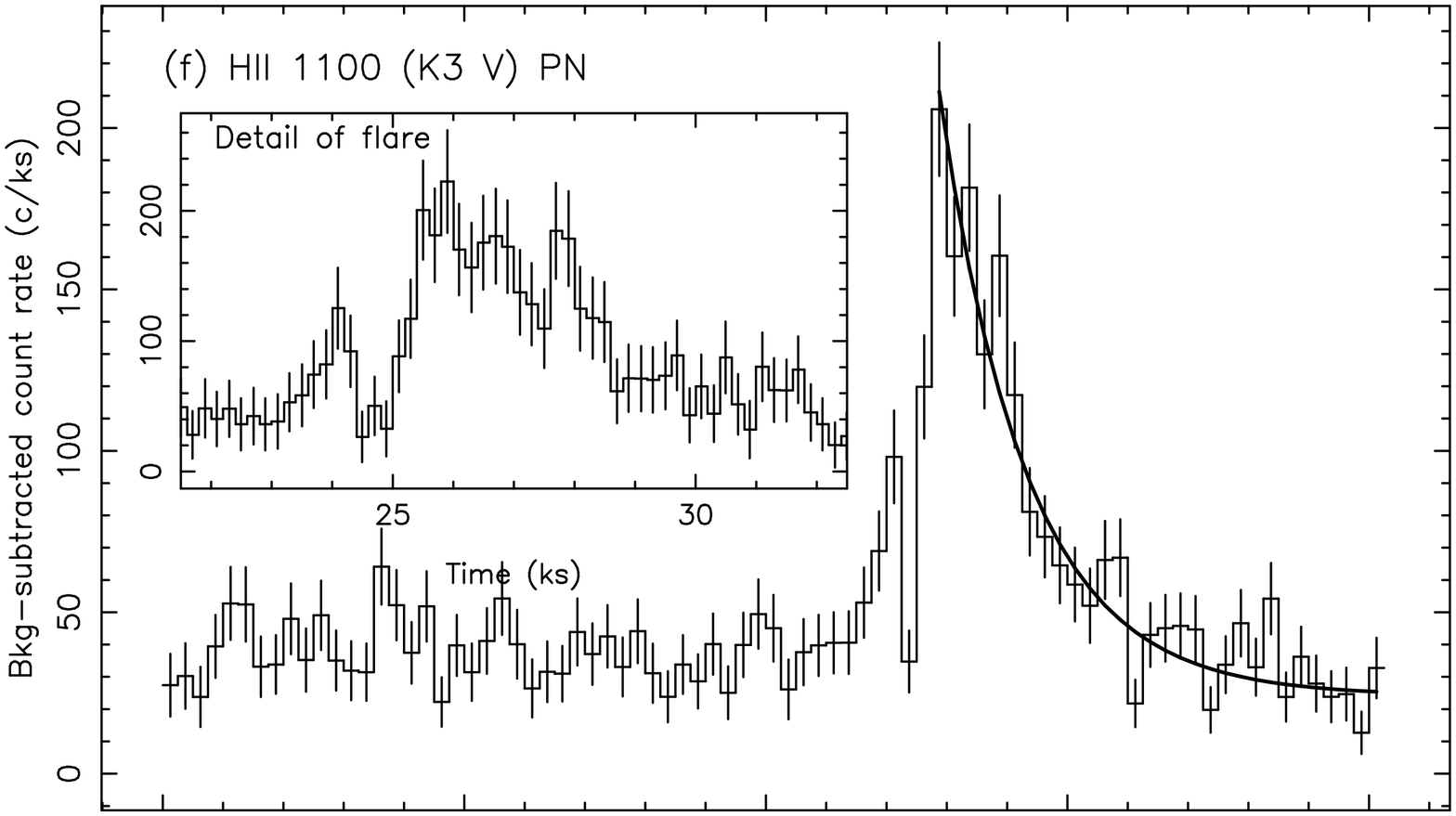}
\end{minipage}\\
\vspace{-0.1cm}
\begin{minipage}[b]{.45\textwidth}
  \centering
  \includegraphics[height=7.1cm,angle=270]{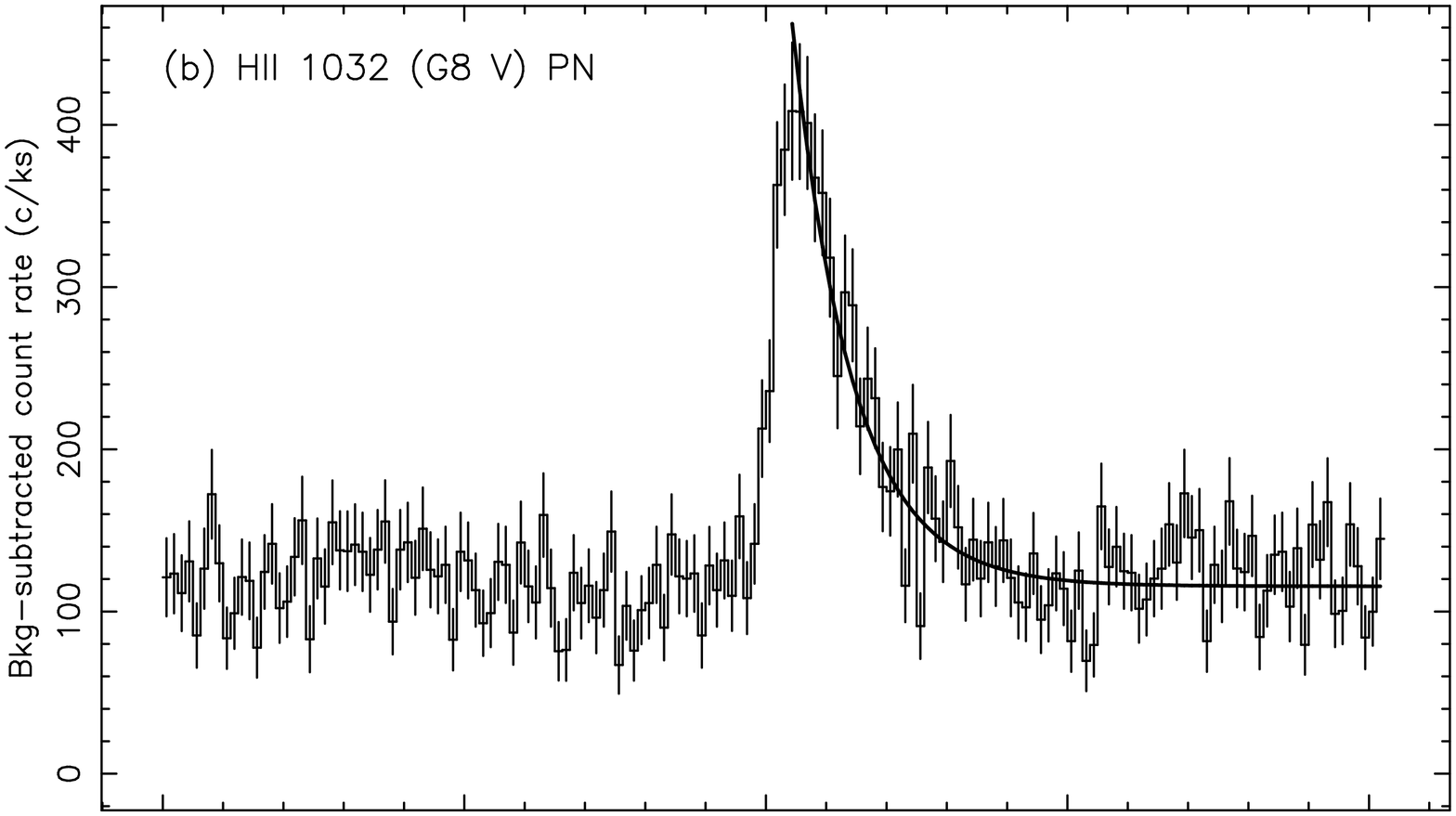}
\end{minipage}%
\begin{minipage}[b]{.45\textwidth}
  \centering
  \includegraphics[height=7.1cm,angle=270]{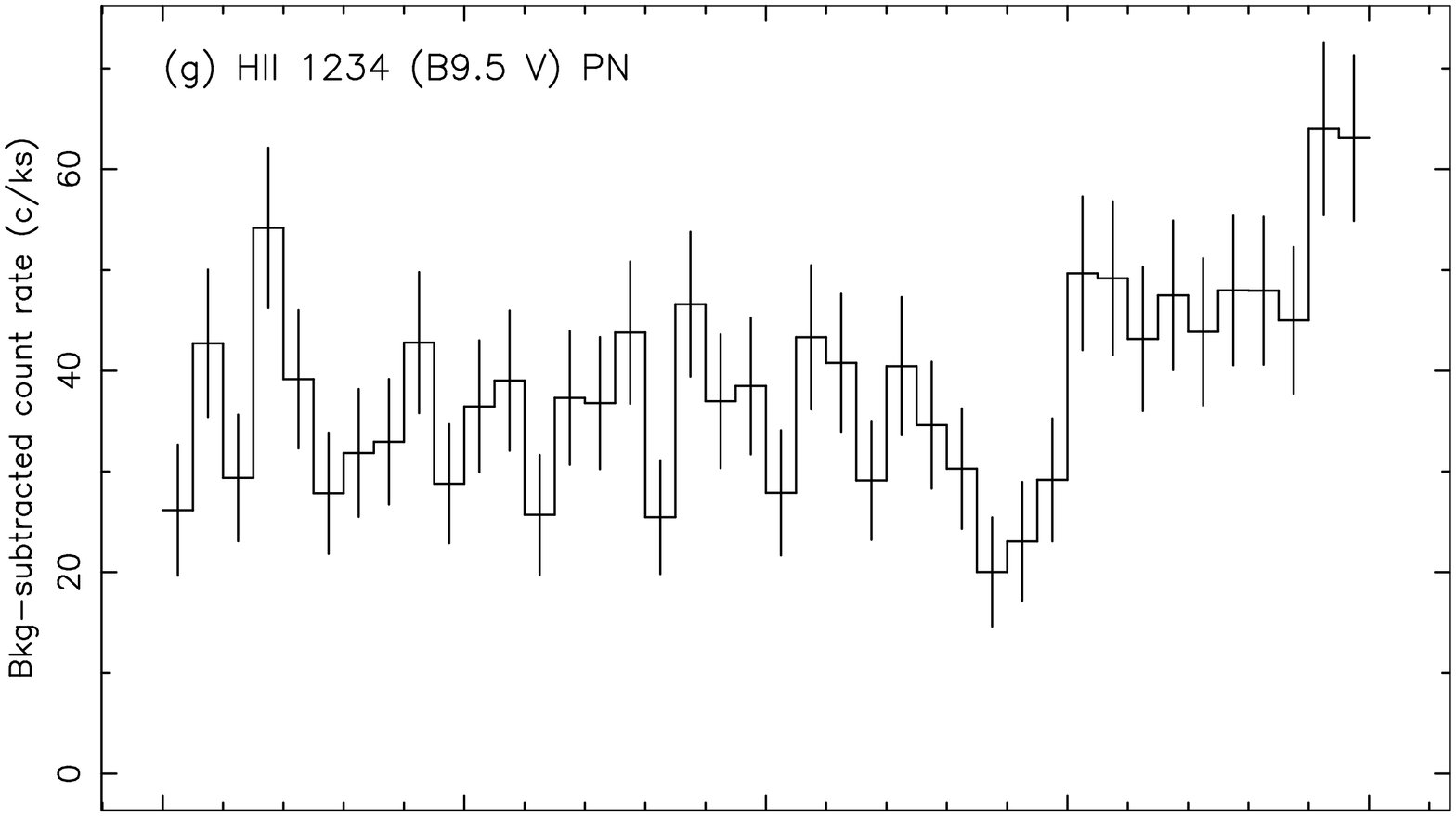}
\end{minipage}\\
\vspace{-0.1cm}
\begin{minipage}[b]{.45\textwidth}
  \centering
  \includegraphics[height=7.1cm,angle=270]{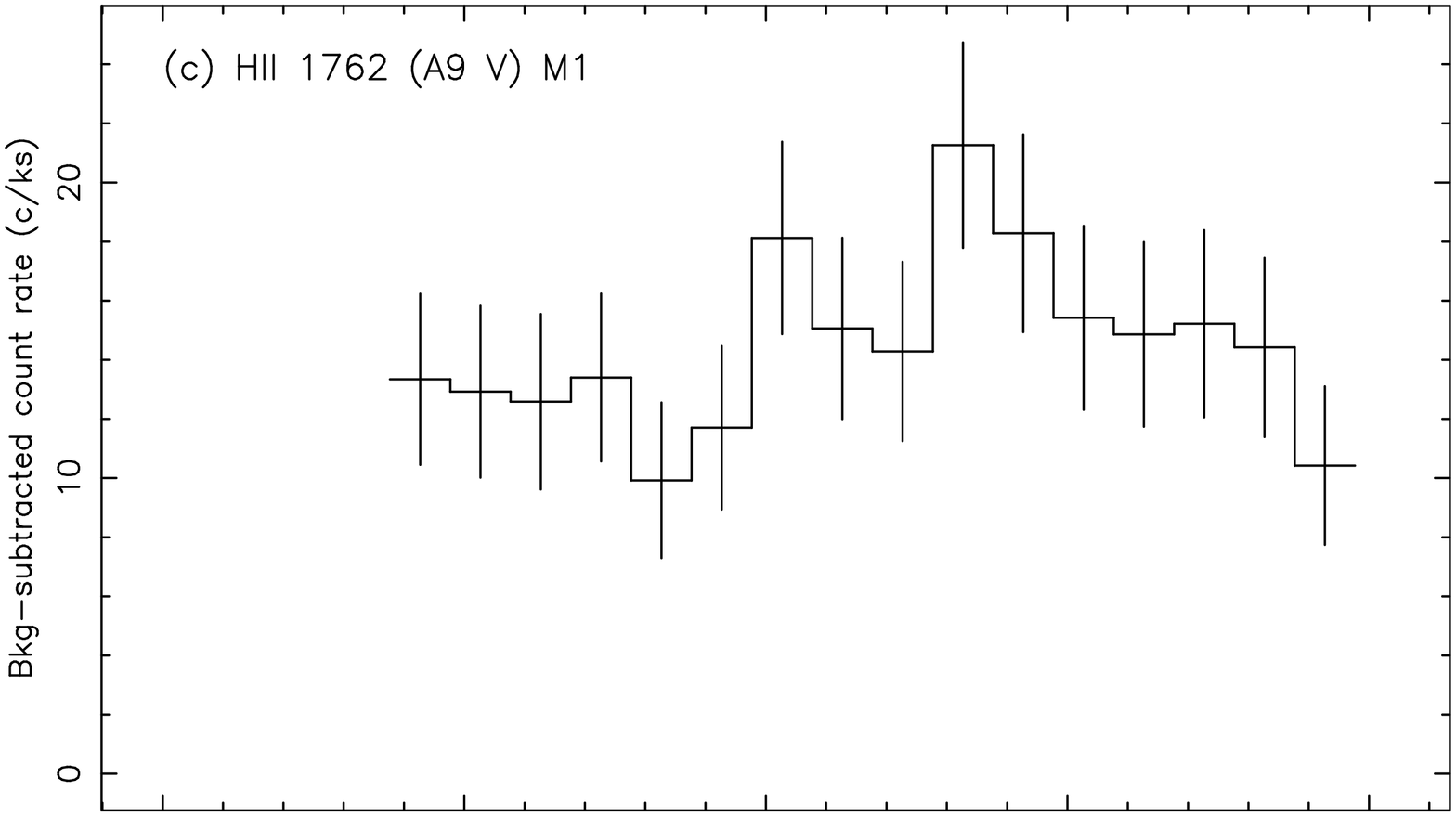}
\end{minipage}%
\begin{minipage}[b]{.45\textwidth}
  \centering
  \includegraphics[height=7.1cm,angle=270]{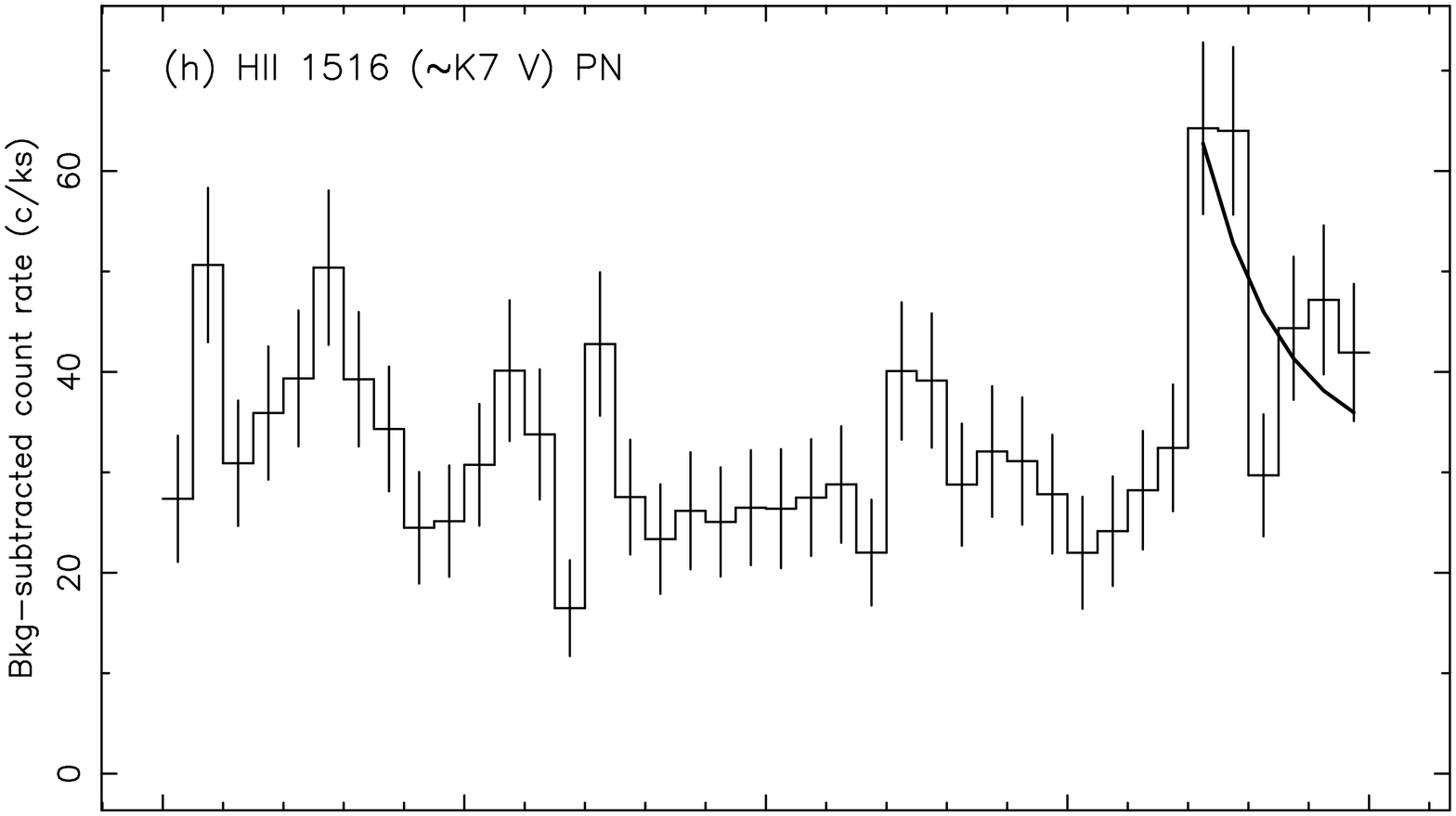}
\end{minipage}\\
\vspace{-0.1cm}
\begin{minipage}[b]{.45\textwidth}
  \centering
  \includegraphics[height=7.1cm,angle=270]{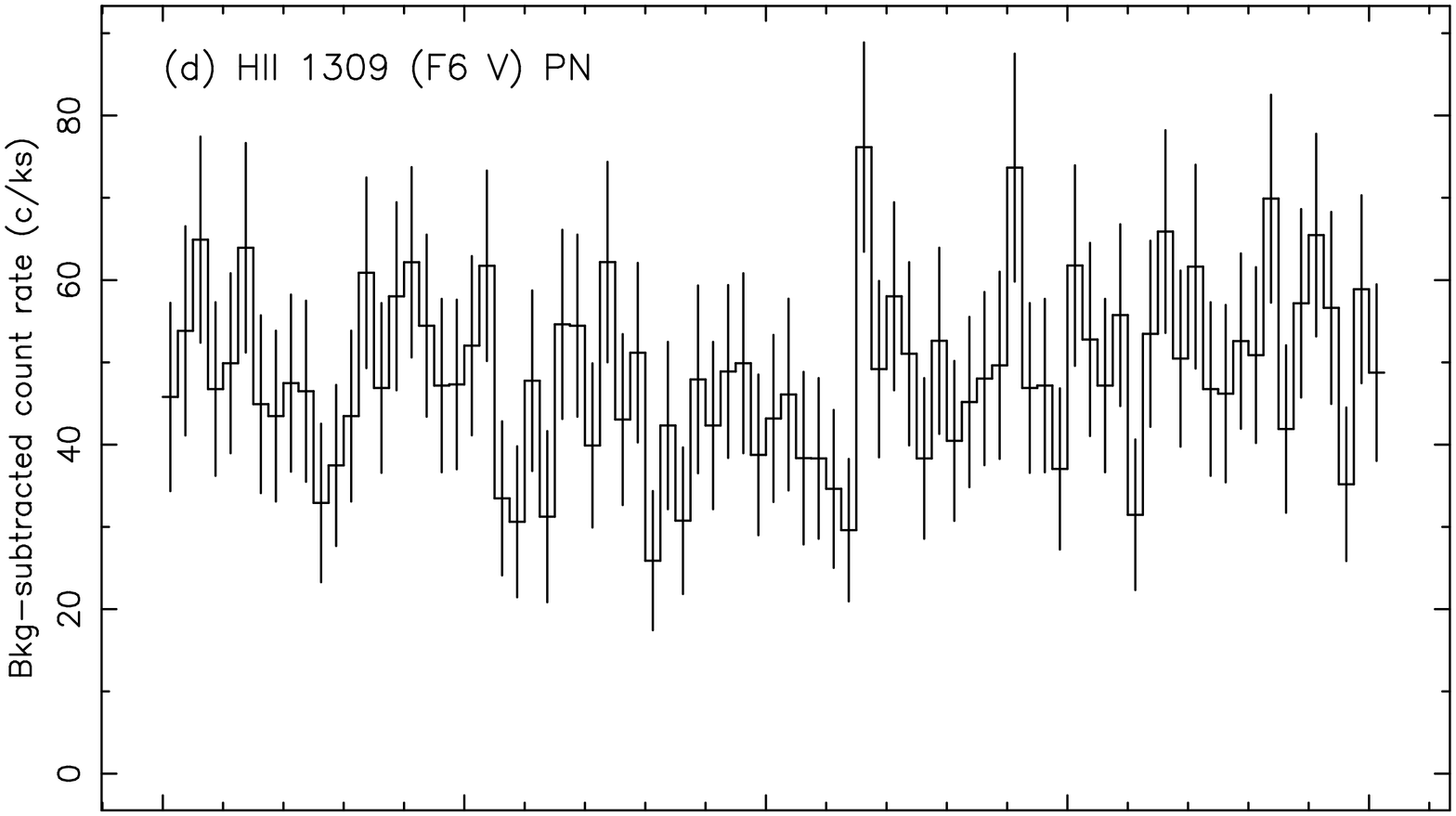}
\end{minipage}%
\begin{minipage}[b]{.45\textwidth}
  \centering
  \includegraphics[height=7.1cm,angle=270]{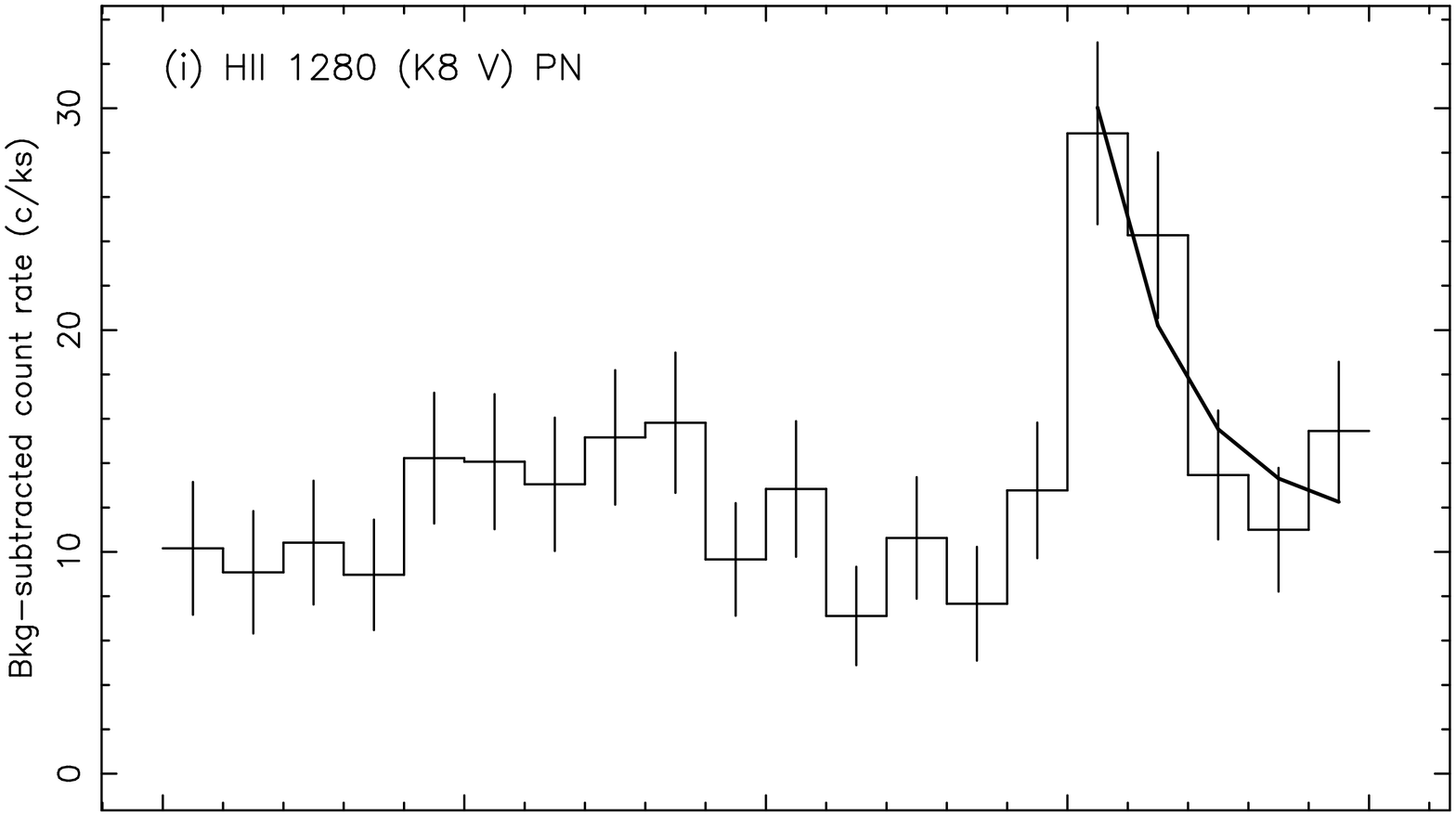}
\end{minipage}\\
\vspace{-0.1cm}
\begin{minipage}[b]{.45\textwidth}
  \centering
  \includegraphics[height=7.1cm,angle=270]{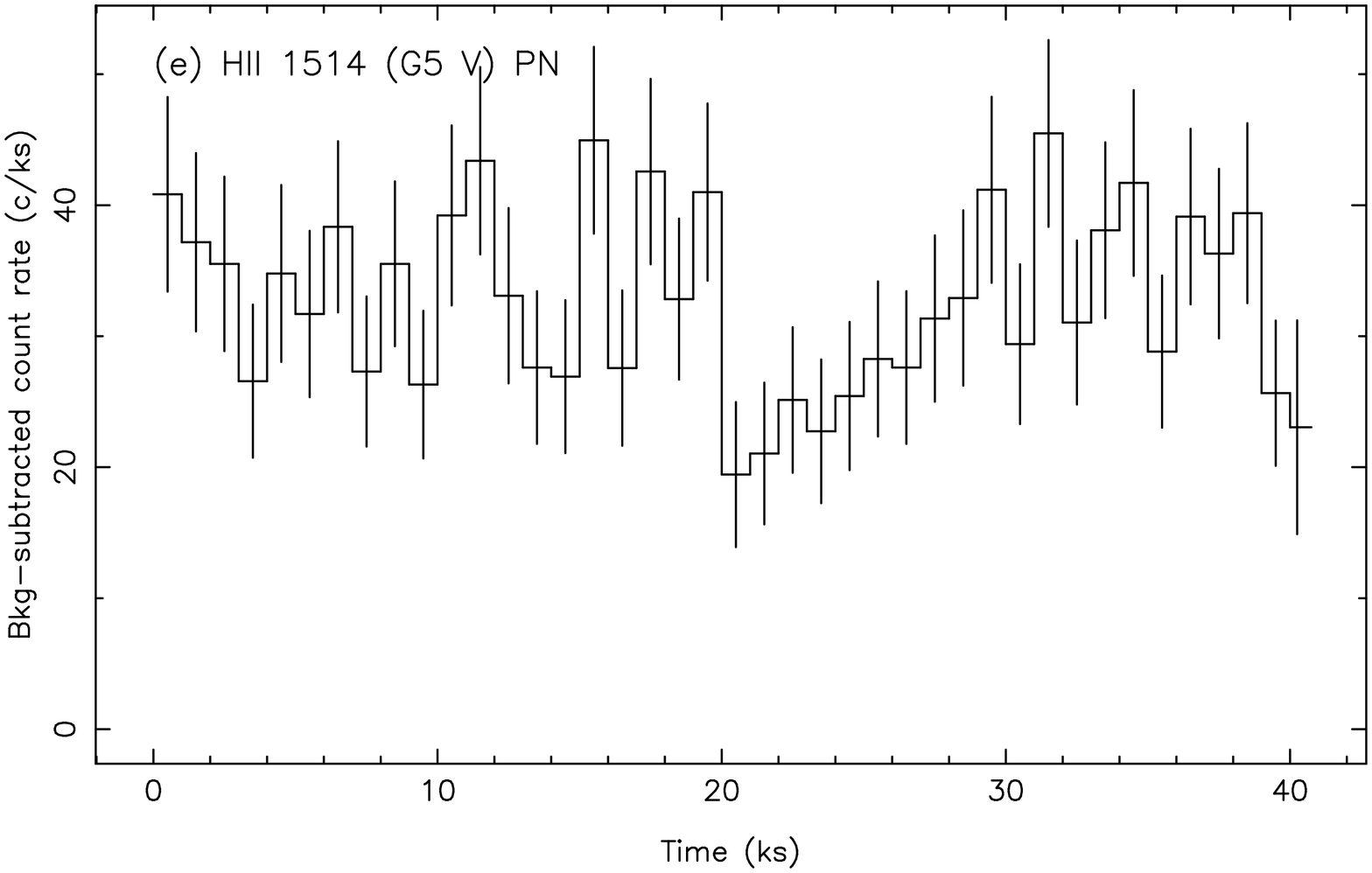}
\end{minipage}%
\begin{minipage}[b]{.45\textwidth}
  \centering
  \includegraphics[height=7.1cm,angle=270]{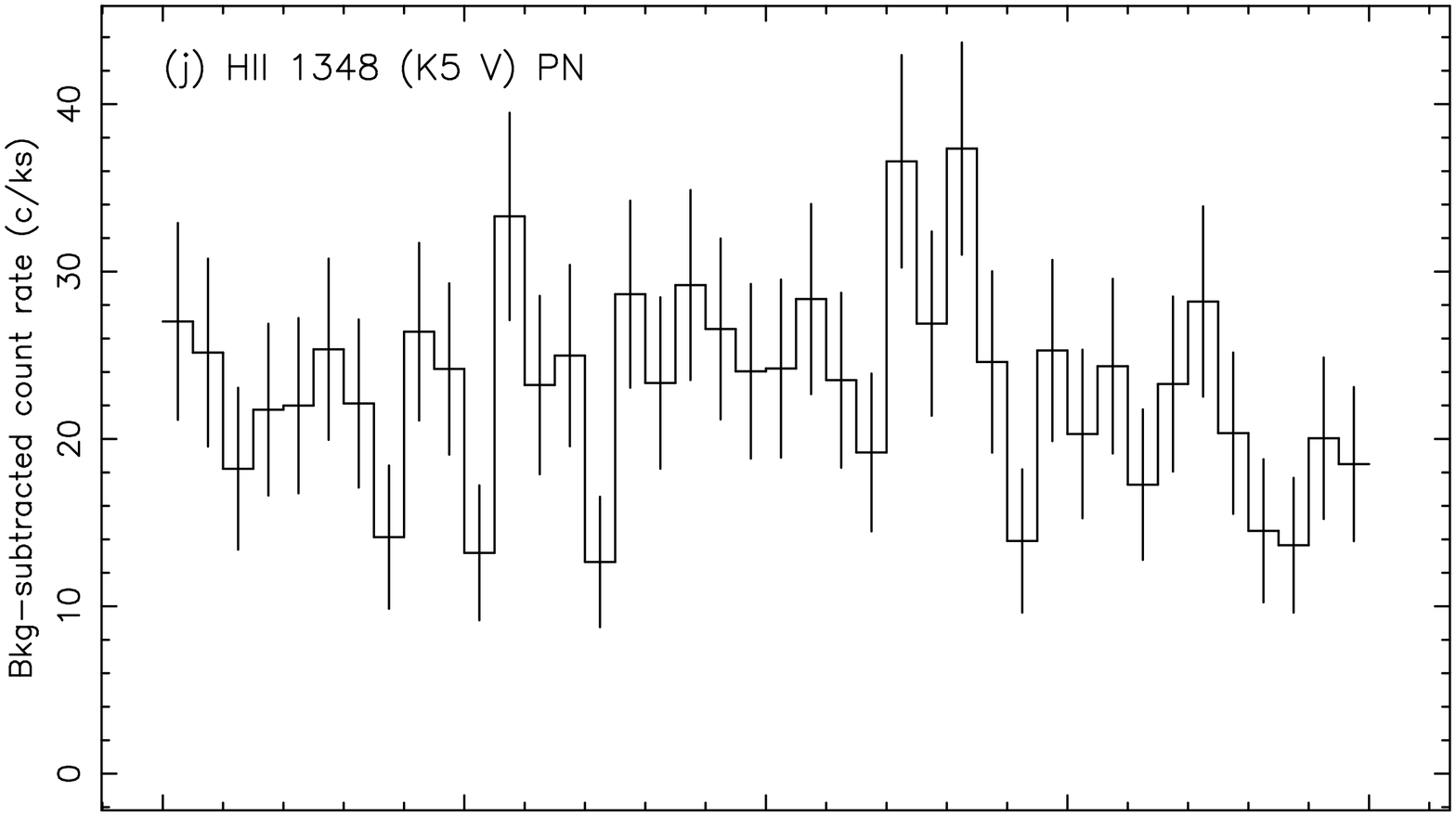}
\end{minipage}\\
\vspace{-0.45cm}
\begin{minipage}[t]{.45\textwidth}
  \centering
\vspace{0.45cm}
\caption{Lightcurves of X-ray-bright Pleiads in the \xmm{} EPIC
field. Each lightcurve is background-subtracted but not
corrected for vignetting or enclosed energy, covers the same time
range and is labelled with the source name, its spectral type, and the
instrument(s) used. For the flaring sources (b, f, h, i) the best-fitting
exponential decay is shown. The columns separate the \emph{sources} by
spectral type (as catalogued for the solar-like; as proposed in
\S~\protect{\ref{sec_res_early}} for the intermediate-type Pleiads): F
and G on the left; K on the right. The sources are ordered by \quies{}
\lx, decreasing from top to bottom.
}
\label{fig_lc}
\end{minipage}%
\begin{minipage}[t]{.45\textwidth}
  \centering
  \includegraphics[height=7.1cm,angle=270]{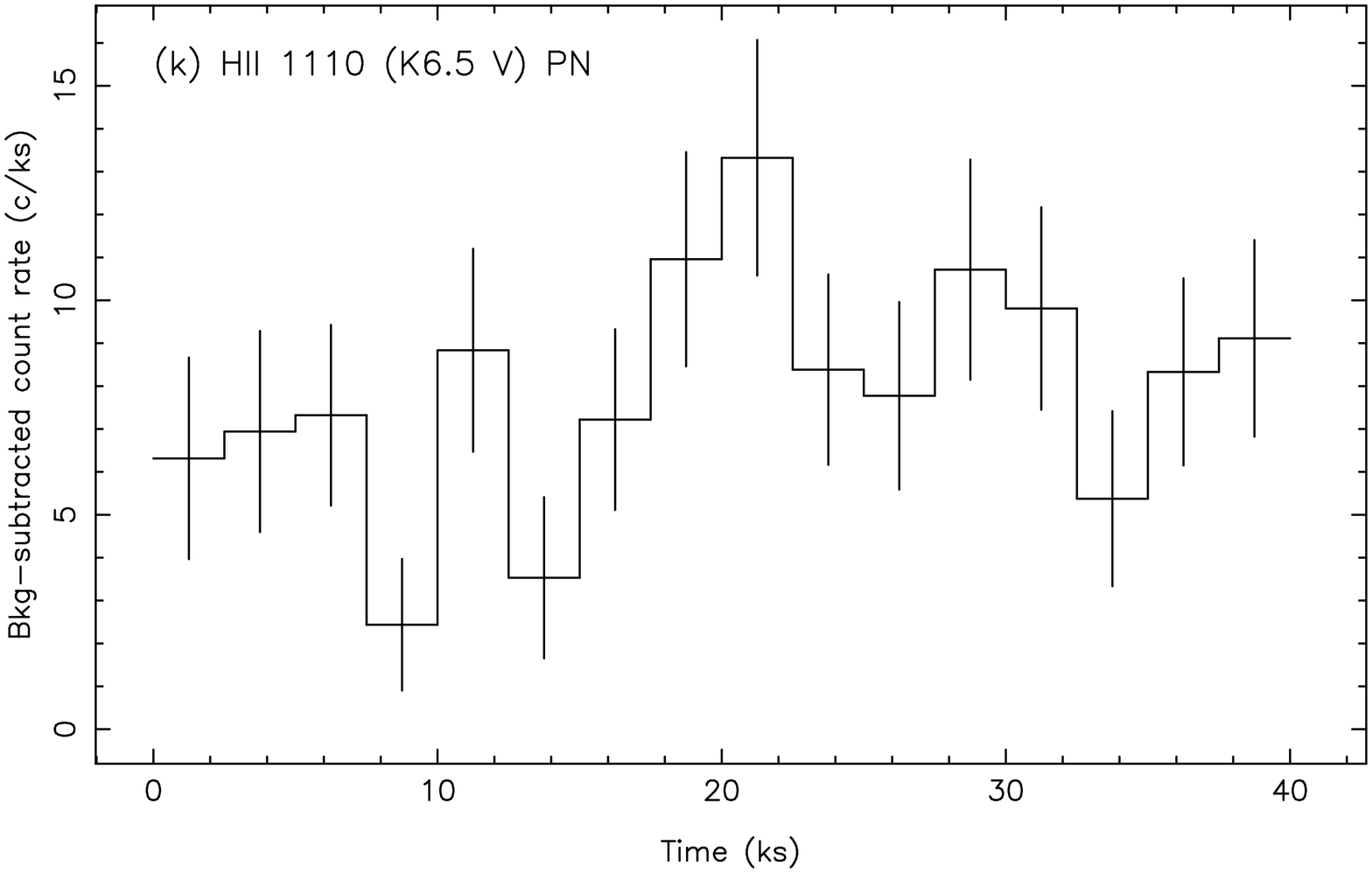}
\end{minipage}\\
}
\end{figure*}

Source and background lightcurves were extracted (from
regions described above) in the 0.3--4.5 keV band and binned to ensure an
average of $> 20$ net source counts per bin, to enable good approximation of
the error on the total counts in each bin, $N$, as
$\sqrt{N}$. Background-subtracted source lightcurves are displayed in
Fig.~\ref{fig_lc}. 

\begin{table}
\centering{
%\begin{minipage}{0.45\textwidth}
\caption{Results of time-series analysis of X-ray-bright Pleiads in
the \xmm{} EPIC field. Columns show: 
(2) instrument(s) used;
(3) radius of extraction region for time-series and spectrum;
(4) bin-size in time-series in s; 
(5) \chisq{} probability of consistency with constant source emission.
For flare-like variability:
(6) peak \loglx{} in the 0.3--4.5 keV band in \ergs; 
(7) exponential decay time-scale (and 90 per cent confidence
interval) in ks.
}
%\vspace{0.2cm}
\label{tbl_var}
\scriptsize
\begin{tabular}{lccrcll}
\hline
\multicolumn{1}{c}{HII}& 
\multicolumn{1}{c}{Ins}& 
\multicolumn{1}{c}{r}&
\multicolumn{1}{c}{$\Delta t$}& 
\multicolumn{1}{c}{$P(\chi^2)$}& 
\multicolumn{1}{c}{[$L_{\rm X}$]$_{\rm pk}$}&
\multicolumn{1}{c}{$\tau_{\rm LC}$ (90\%)}\\
\multicolumn{1}{c}{(1)} & 
\multicolumn{1}{c}{(2)} & 
\multicolumn{1}{c}{(3)} & 
\multicolumn{1}{c}{(4)} & 
\multicolumn{1}{c}{(5)} & 
\multicolumn{1}{c}{(6)} & 
\multicolumn{1}{c}{(7)} \\
\hline
1234 & \pn & 50 & 1000 & 0.00 &  \\
1384 & \pn & 60 &  500 & 0.04 & \\
     & M1 & 60 &  500 & 0.41 & \\
     & M2 & 60 &  500 & 0.14 & \\
     &M1+2& 60 &  250 & 0.37 & \\
1762 & M1 & 50 & 2000 & 0.57 & \\
1309 & \pn & 50 &  500 & 0.63 & \\
1514 & \pn & 40 & 1000 & 0.17 & \\
1032 & \pn & 60 &  250 & 0.00 & 30.8 &2.0 (1.7--2.3)\\
1100 & \pn & 50 &  500 & 0.00 & 30.3 &2.9 (2.6--3.3)\\
1348 & \pn & 30 & 1000 & 0.07 &\\	         
1110 & \pn & 60 & 2500 & 0.04 &\\	         
1516 & \pn & 30 & 1000 & 0.00 & 29.7 &2.6 (1.2--6.8)\\
1280 & \pn & 40 & 2000 & 0.00 & 29.6 &2.7 (1.5--4.7)\\
\hline
\end{tabular}
%\end{minipage}
}
\end{table}

We visually identified flare-like behaviour, defined as an isolated
event of heightened emission peaking at factor $> 2$ above the mean
``\quies'' level outside that event. The decay time-scale of each flare
was estimated by fitting an exponential decay to the lightcurve,
fixing the time of peak emission to the bin containing the peak number
of counts. The peak and \quies{} count-rates and the $e$-folding
decay time-scale were fitted as free parameters. General variability
was assessed by applying a \chisq{} test 
against constancy to the background-subtracted source lightcurve.
While the \chisq{} test is flawed in that its result depends on the
binning of data, we feel it is more robust than the Kolmogorov--Smirnov
(K--S) test, which is applied to unbinned timing data but is not
strictly applicable in the presence of significant varying background,
which is the case here. Flare parameters and results of the \chisq{}
test are given in Table~\ref{tbl_var}.

%%%%%%%%%%%%%%%%%%%%%%%%%%%%%%%%%%%%%%%%%%%%%%%%%%%%%%%%%%%%%%%%%%%%%%%%%%%%%%%
% Table: Spectral fitting parameters - Z
%%%%%%%%%%%%%%%%%%%%%%%%%%%%%%%%%%%%%%%%%%%%%%%%%%%%%%%%%%%%%%%%%%%%%%%%%%%%%%%

\begin{table*}
%\centering{
\begin{minipage}{\textwidth}
\caption{Spectral parametrization of \quies{} emission from
X-ray-bright Pleiads in the \xmm{} EPIC field. The fitting procedure
is described in \S~\protect\ref{sec_xdata_spex}. Columns show:
(2) instrument(s) used in fit; 
(3) net source counts in spectrum;
(4) and (5) temperatures in keV of cooler and hotter components respectively;
(6) $\log$ emission measure of cooler component in cm$^{-3}$;
(7) ratio of emission measures of hotter to cooler component;
(8) metallicity;
(9) \loglx{} in \ergs{} in 0.3--4.5 keV band; 
(10) \loglxlbol;
(11) \chisq{} / degrees of freedom and null-hypothesis probability. 
Columns (4), (5), (7), and (8) give the best-fitting
value and (90 per cent confidence interval). HII~1384's point spread
function is only partially included on the \pn{} CCD, so no values are
given for columns (6), (9) and (10).
}
%\vspace{0.2cm}
\label{tbl_spec}
\scriptsize
\begin{tabular}{lcrllcllccr}
\hline
\multicolumn{1}{c}{HII} & 
\multicolumn{1}{c}{Ins} & 
\multicolumn{1}{c}{$N_{\rm X}$} & 
\multicolumn{1}{c}{$kT_1$ (90\%)} &
\multicolumn{1}{c}{$kT_2$ (90\%)} &
\multicolumn{1}{c}{[$EM_1$]} &
%\multicolumn{1}{c}{$EM_2/EM_1$ (90\%)} &
\multicolumn{1}{c}{$EM$ ratio (90\%)} &
\multicolumn{1}{c}{$Z$ (90\%)} &
\multicolumn{1}{c}{[$L_{\rm X}$]} &
\multicolumn{1}{c}{[$L_{\rm X}/L_{\rm bol}$]} &
\multicolumn{1}{c}{\chisq/$\nu$ ($P$)}\\
\multicolumn{1}{c}{(1)} & 
\multicolumn{1}{c}{(2)} & 
\multicolumn{1}{c}{(3)} & 
\multicolumn{1}{c}{(4)} & 
\multicolumn{1}{c}{(5)} & 
\multicolumn{1}{c}{(6)} & 
\multicolumn{1}{c}{(7)} & 
\multicolumn{1}{c}{(8)} & 
\multicolumn{1}{c}{(9)} & 
\multicolumn{1}{c}{(10)} & 
\multicolumn{1}{c}{(11)}\\
\hline
%HII	In N_sp    kT1  (90%)          kT2  (90%)          EM1     EM2/EM1 (90%)       Z    (90%)          Lx_sp  Lx/Lbol_sp chi/dof (Prob)\\
%
1234  & \pn & 1051 & 0.37 (0.31--0.50) & 1.08 (0.98--1.31) & 52.18 & 1.49 (1.08--2.06) & 0.26 (0.17--0.41) & 29.53 & $-5.62$ &  39/ 49 (0.86)\\
1384  & \pn & 1908 & 0.50 (0.37--0.60) & 1.02 (0.91--1.18) & --    & 0.84 (0.41--1.27) & 0.26 (0.20--0.36) &  --   &  --     &  79/ 79 (0.49)\\
      &  M1 & 1443 & 0.54 (0.45--0.63) & 1.01 (0.88--1.27) & 52.93 & 0.80 (0.30--1.22) & 0.26 (0.19--0.37) & 30.17 & $-4.53$ &  42/ 50 (0.77)\\
      &  M2 & 1493 & 0.61 (0.49--0.66) & 1.26 (0.94--1.94) & 53.15 & 0.43 (0.18--1.02) & 0.18 (0.13--0.26) & 30.21 & $-4.49$ &  42/ 53 (0.87)\\
      & M1,2& 2936 & 0.53 (0.49--0.58) & 1.01 (0.95--1.35) & 52.99 & 0.88 (0.42--1.06) & 0.22 (0.17--0.27) & 30.20 & $-4.50$ &  90/107 (0.88)\\
1762  &  M1 &  485 & 0.64 (0.61--0.68) &                   & 52.19 &                   & 1.00              & 29.63 & $-4.80$ &  14/ 20 (0.82)\\
1309  & \pn & 1982 & 0.56 (0.54--0.59) &                   & 52.26 &                   & 0.54 (0.38--0.79) & 29.48 & $-4.49$ &  77/ 86 (0.74)\\
1514  & \pn & 1347 & 0.56 (0.53--0.59) &                   & 52.29 &                   & 0.29 (0.23--0.47) & 29.31 & $-4.28$ &  71/ 60 (0.15)\\
1032  & \pn & 3674 & 0.62 (0.58--0.67) & 1.23 (1.08--1.35) & 52.72 & 1.48 (1.10--1.93) & 0.33 (0.26--0.42) & 30.14 & $-3.14$ & 149/145 (0.40)\\
1100  & \pn &  906 & 0.37 (0.30--0.47) & 0.93 (0.83--1.09) & 52.35 & 1.11 (0.80--1.80) & 0.20 (0.13--0.50) & 29.55 & $-3.48$ &  37/ 41 (0.64)\\
1348  & \pn &  942 & 0.34 (0.27--0.45) & 0.94 (0.77--1.09) & 52.03 & 0.78 (0.52--1.38) & 0.18 (0.10--0.32) & 29.12 & $-3.77$ &  35/ 43 (0.79)\\
1110  & \pn &  322 & 0.34 (0.28--0.42) & 1.02 (0.79--1.29) & 51.36 & 0.94 (0.58--1.47) & 1.00              & 29.00 & $-3.69$ &  17/ 18 (0.53)\\
1516  & \pn & 1036 & 0.32 (0.25--0.40) & 0.92 (0.78--1.04) & 52.14 & 1.19 (0.87--2.44) & 0.22 (0.13--0.36) & 29.36 & $-3.11$ &  56/ 48 (0.20)\\
1280  & \pn &  432 & 0.32 (0.27--0.37) & 1.32 (1.06--1.74) & 51.56 & 1.39 (0.96--1.98) & 1.00              & 29.22 & $-3.09$ &  17/ 22 (0.76)\\
\hline
\end{tabular}
\vspace{-0.3cm}
\end{minipage}
%}
\end{table*}

\subsection{Spectral analysis}
\label{sec_xdata_spex}

In the extraction of spectra of variable sources, periods affected by
clear flares (e.g. HII~1516) or non-flare-like high emission
(e.g. HII~1234) were excluded to enable analysis of ``\quies''
emission, i.e. not strongly biased by a single transient event. Where
sufficient 
counts were available, spectra from these periods were analysed separately. 
For each spectrum, an ancilliary response file (ARF) was generated to
account for bad pixels and chip gaps, the spatial variations of
effective area and quantum efficiency, vignetting, and losses due to
the finite source extraction area. Finally, each source spectrum was
grouped to a minimum of 20 counts per bin. All data bins containing fewer than
20 counts, and all data outside the range 0.3--4.5 keV, were ignored
in spectral analysis. The latest (as of October 2002) \pn{} and MOS
redistribution matrices were used, with the ARFs, to model the
instrumental responses. 

The X-ray emission of a stellar corona is due to hot,
optically thin plasma, and the spectrum is expected to be a superposition of
the spectra of many individual loops of hot plasma at different temperatures.
Hence, the form of the differential emission measure (DEM) as a
function of temperature may be rather complex. Additionally, 
elemental abundances in coronae may differ from solar abundances.
However, the relatively low energy resolution ($\approx 70$ eV)
provided by EPIC, and signal-to-noise typical of bright coronae at
the distance of the Pleiades, require us to approximate the DEM to
one (1-T) or two (2-T) discrete temperature components, and estimate
abundances using a metallicity parameter, $Z$, in which the abundances 
of individual metals are fixed to the ratios observed in the solar
photosphere but the total abundance of metals relative to that of H
may be fixed to the solar value (\zsol), or fitted freely (\zfree).
The \mekal{} plasma emission code was used, and 
the absorbing column density was fixed to the nominal value for the
Pleiades (see \S~\ref{sec_memb}) of \nh{} $= 2 \times 10^{20}$ cm$^{-2}$, as
exclusion of data below 0.3 keV prevents constraint of \nh{} at
such low values.

\begin{figure*}
\centering{
\begin{minipage}[b]{.45\textwidth}
  \centering
  \includegraphics[height=7.1cm,angle=270]{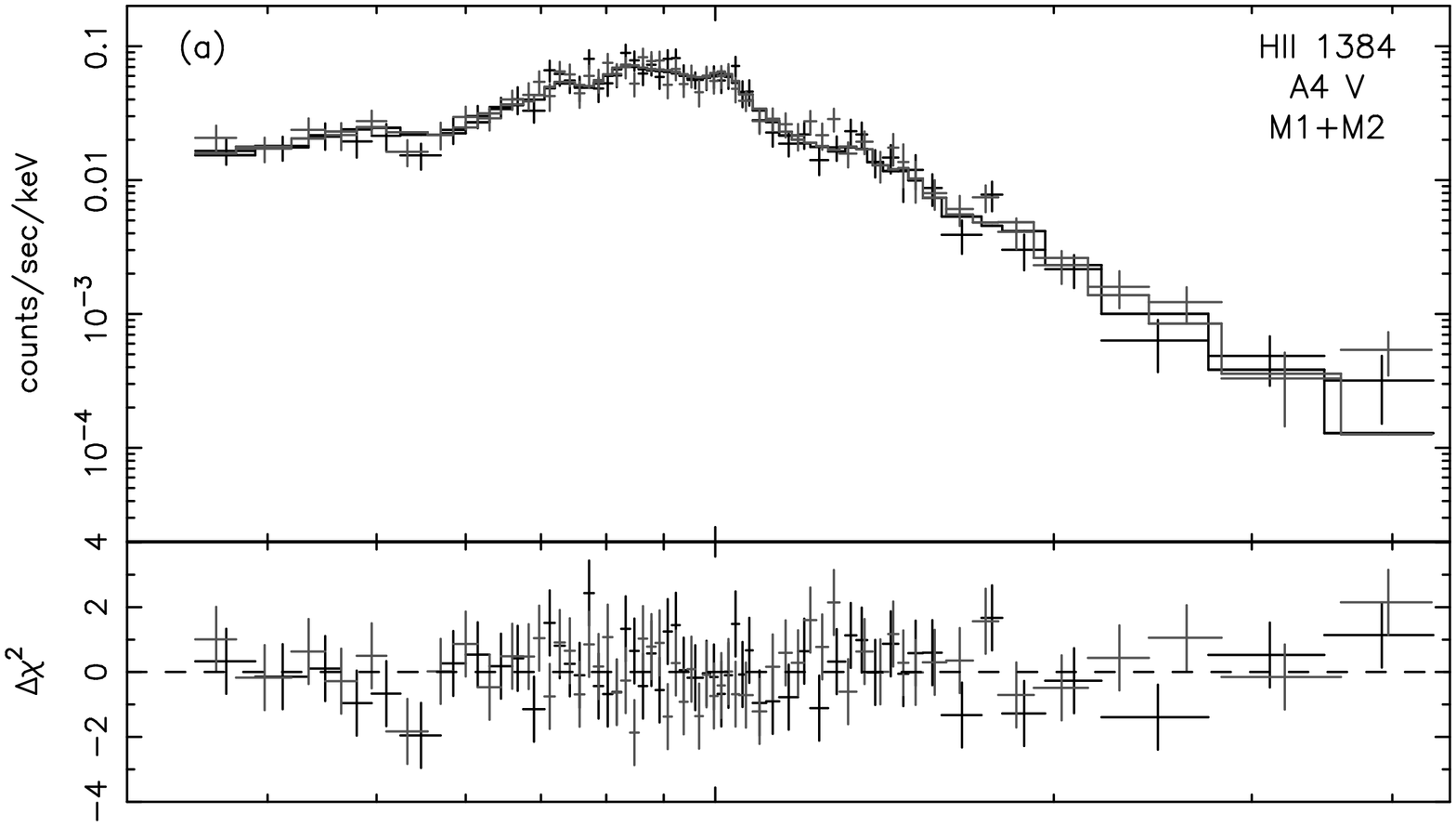}
\end{minipage}%
\begin{minipage}[b]{.45\textwidth}
  \centering
  \includegraphics[height=7.1cm,angle=270]{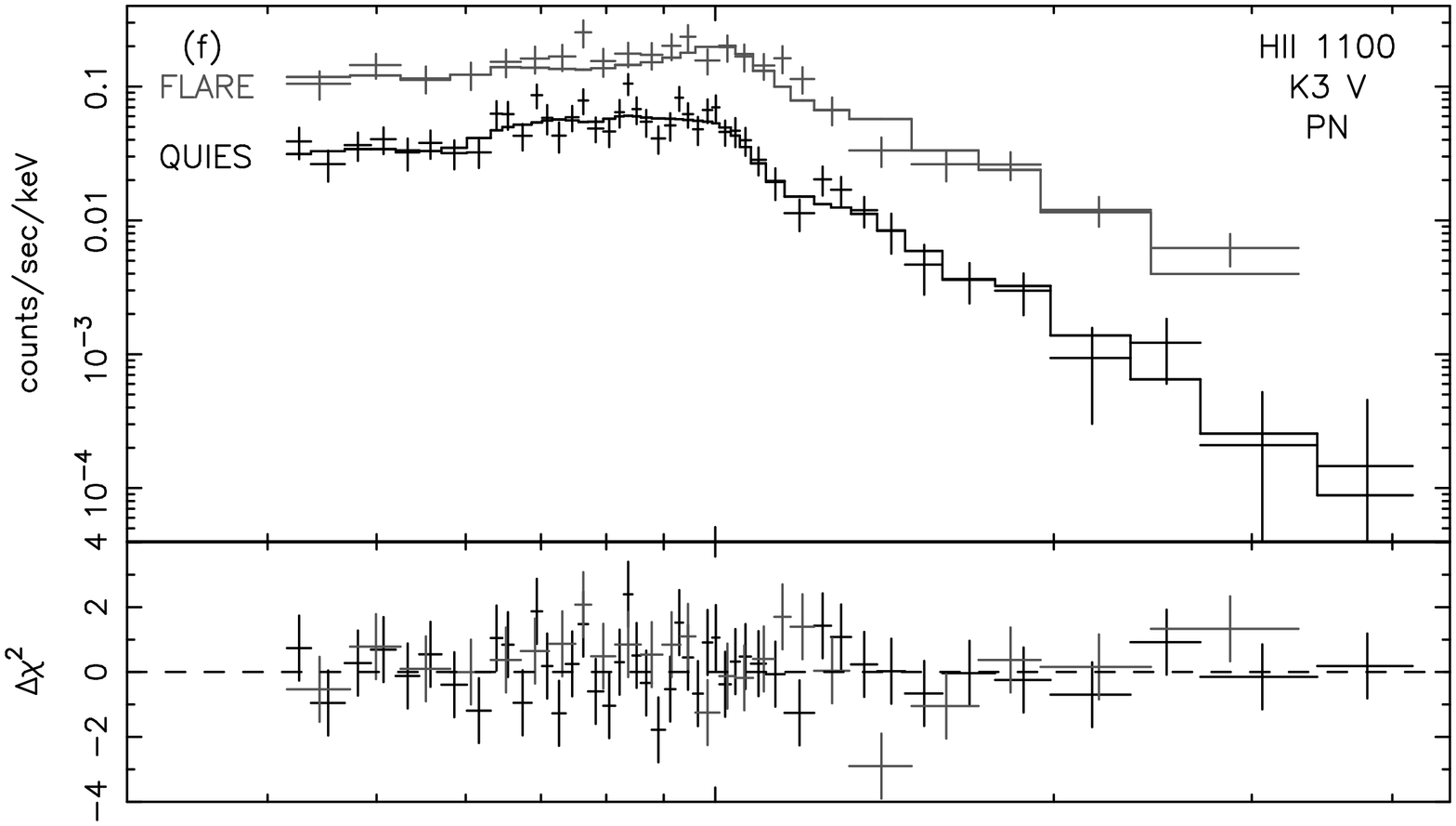}
\end{minipage}\\
\vspace{-0.1cm}
\begin{minipage}[b]{.45\textwidth}
  \centering
  \includegraphics[height=7.1cm,angle=270]{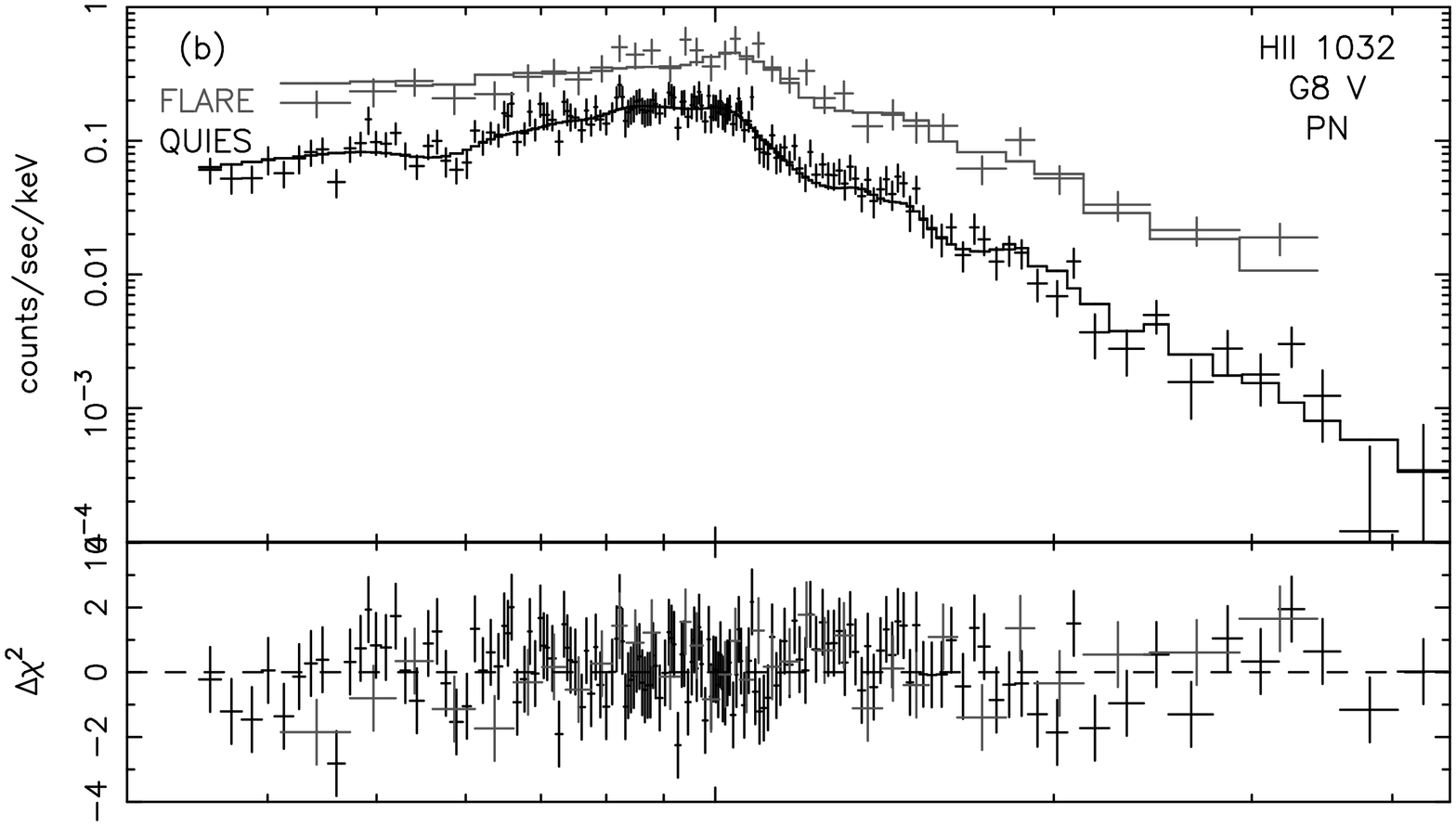}
\end{minipage}%
\begin{minipage}[b]{.45\textwidth}
  \centering
  \includegraphics[height=7.1cm,angle=270]{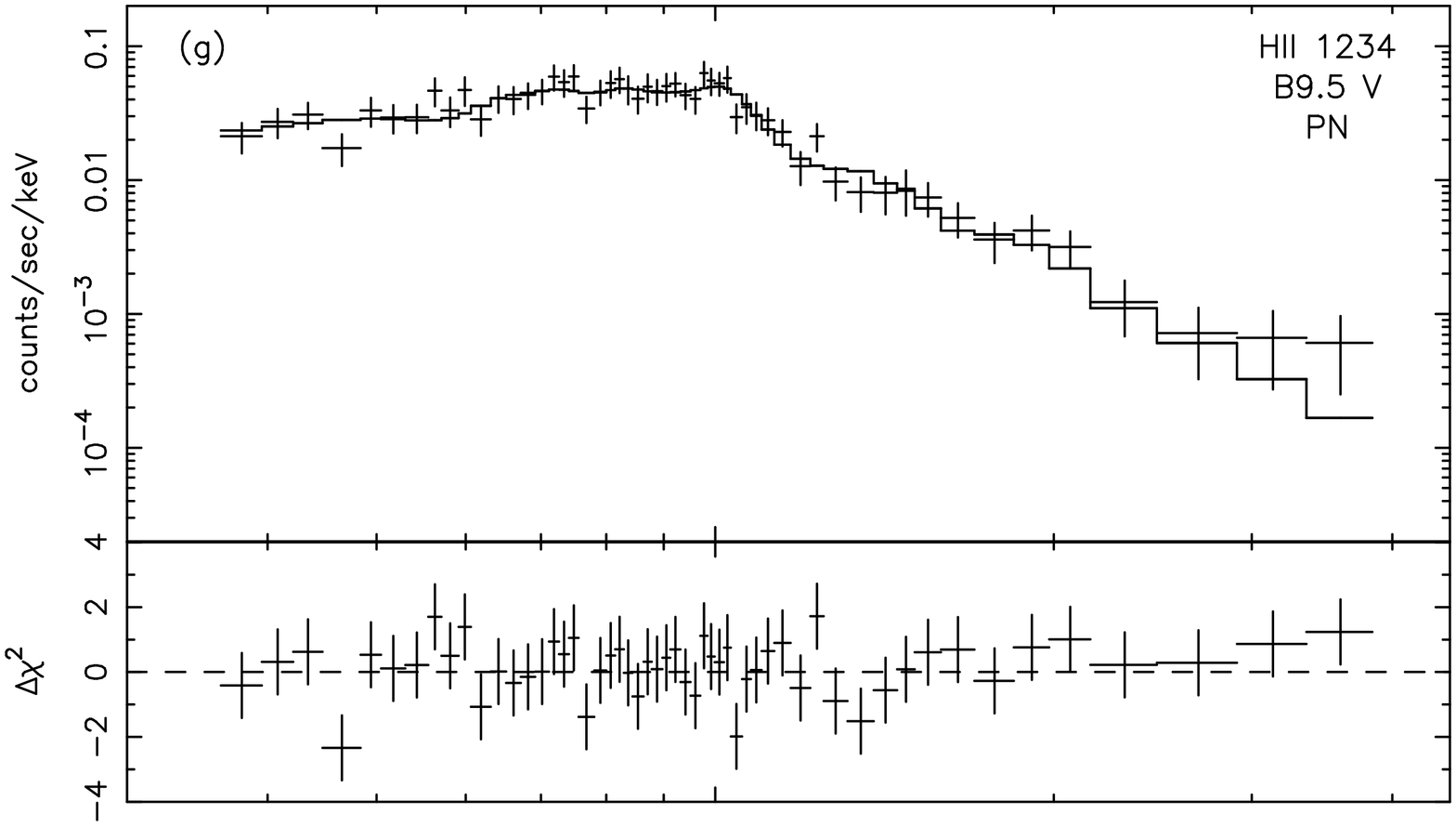}
\end{minipage}\\
\vspace{-0.1cm}
\begin{minipage}[b]{.45\textwidth}
  \centering
  \includegraphics[height=7.1cm,angle=270]{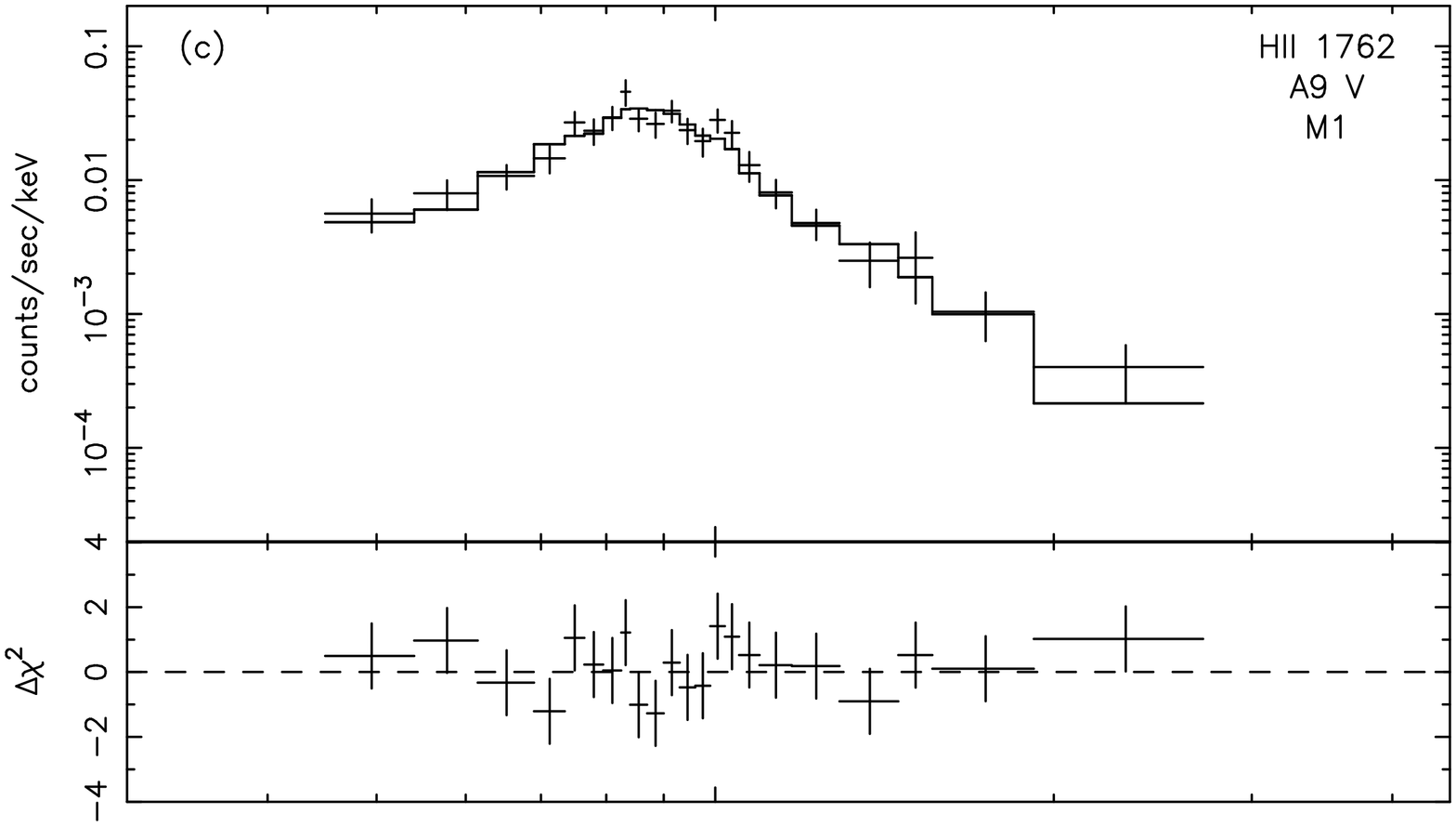}
\end{minipage}%
\begin{minipage}[b]{.45\textwidth}
  \centering
  \includegraphics[height=7.1cm,angle=270]{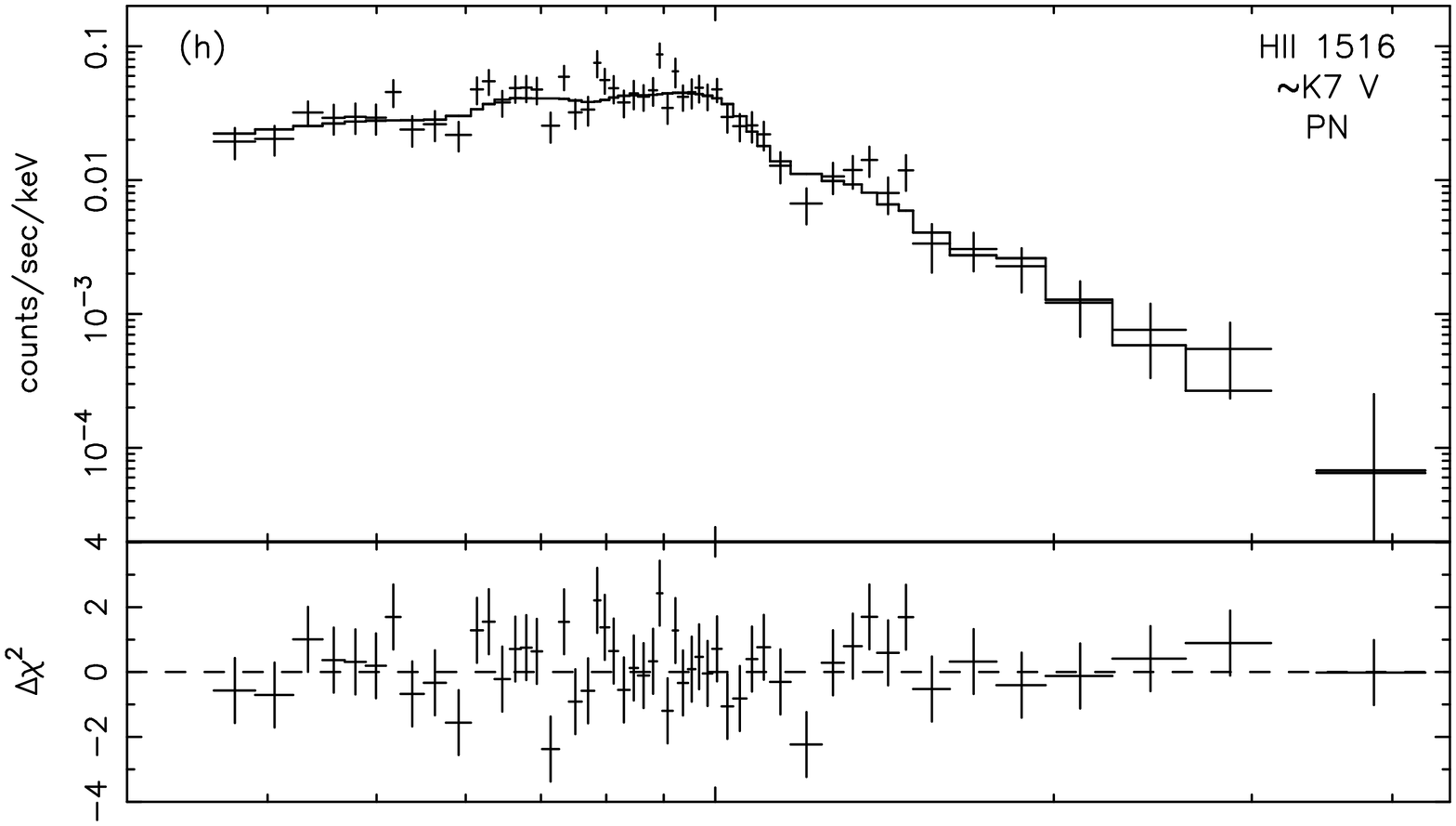}
\end{minipage}\\
\vspace{-0.1cm}
\begin{minipage}[b]{.45\textwidth}
  \centering
  \includegraphics[height=7.1cm,angle=270]{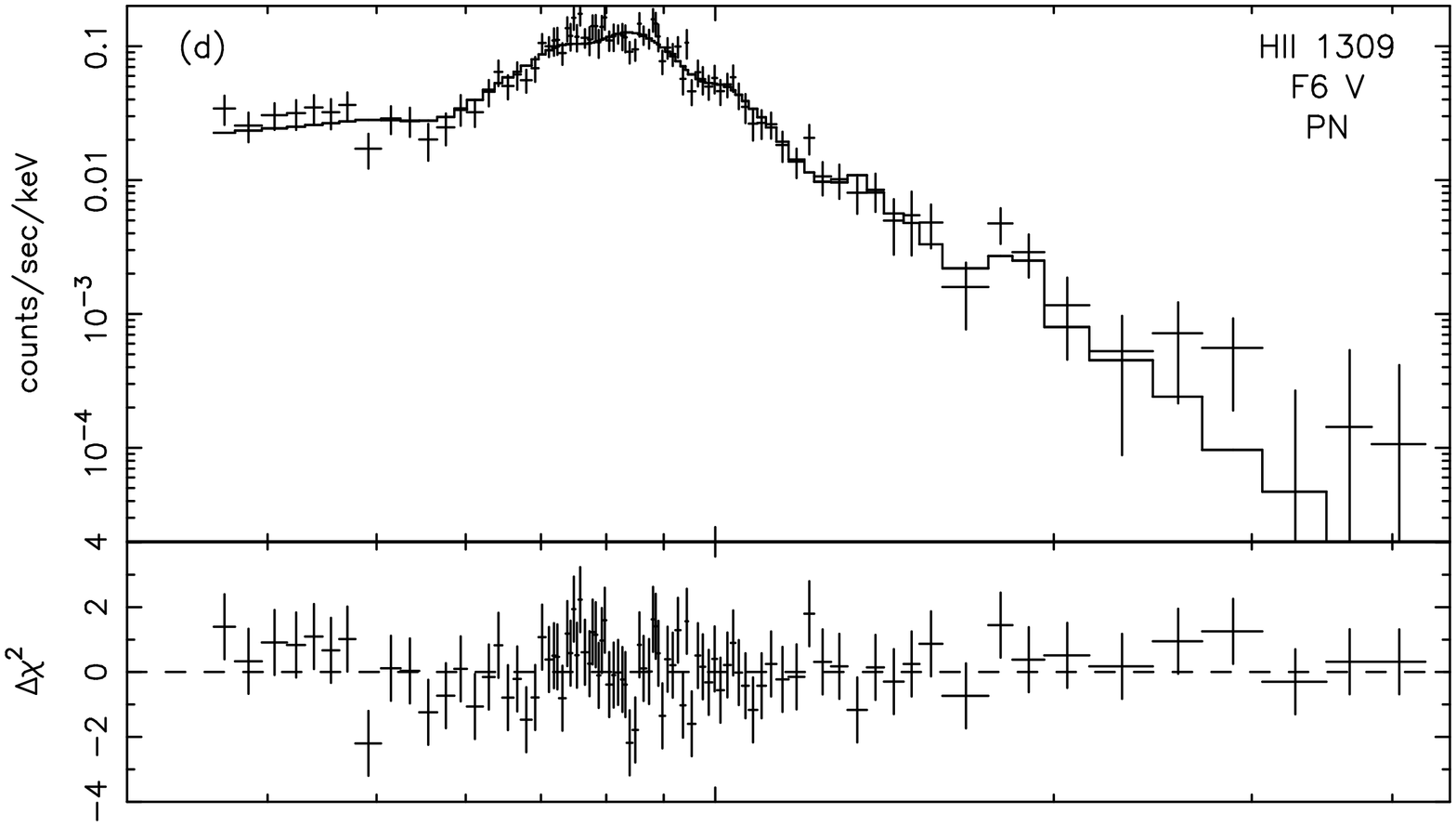}
\end{minipage}%
\begin{minipage}[b]{.45\textwidth}
  \centering
  \includegraphics[height=7.1cm,angle=270]{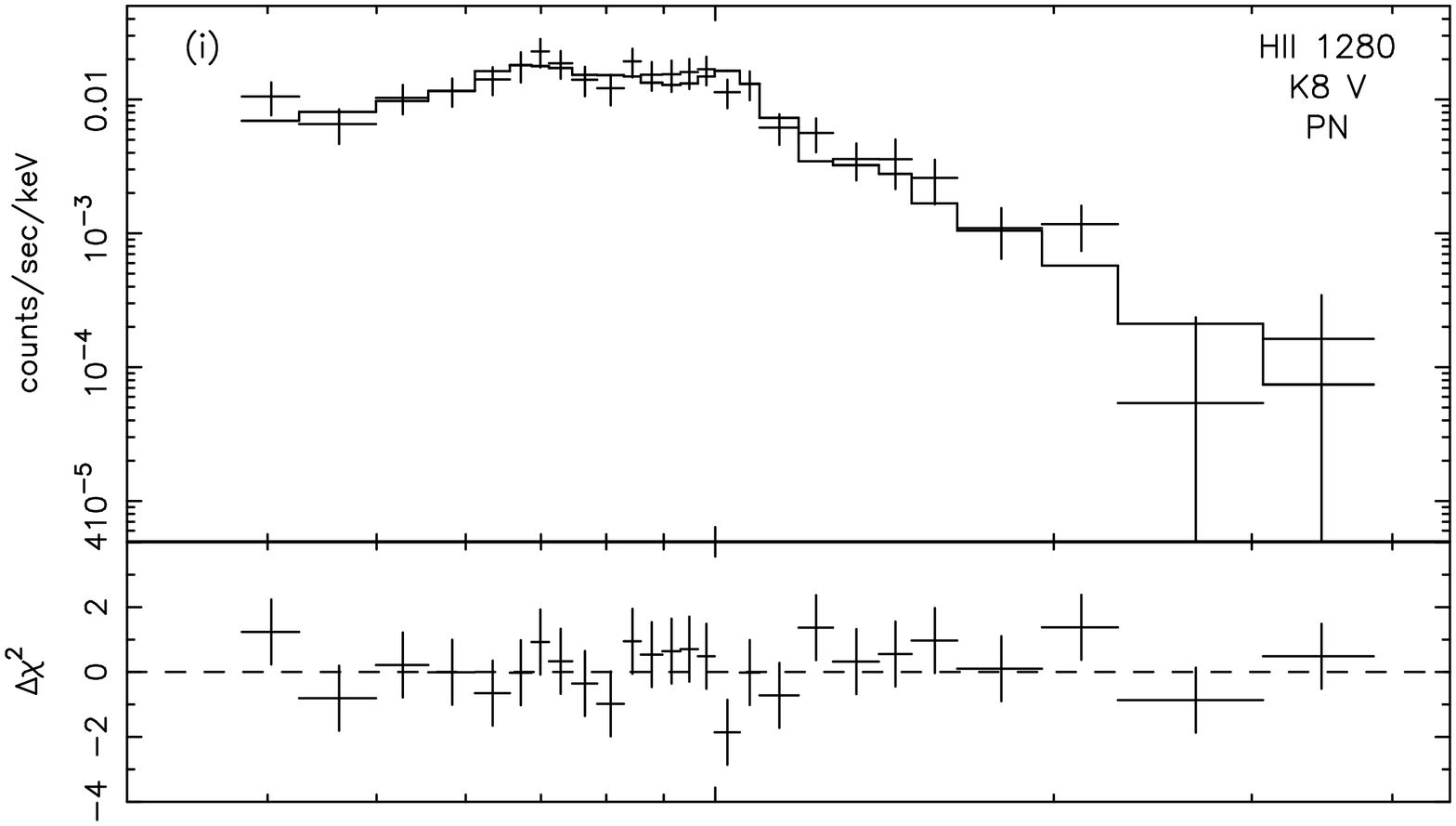}
\end{minipage}\\
\vspace{-0.1cm}
\begin{minipage}[b]{.45\textwidth}
  \centering
  \includegraphics[height=7.1cm,angle=270]{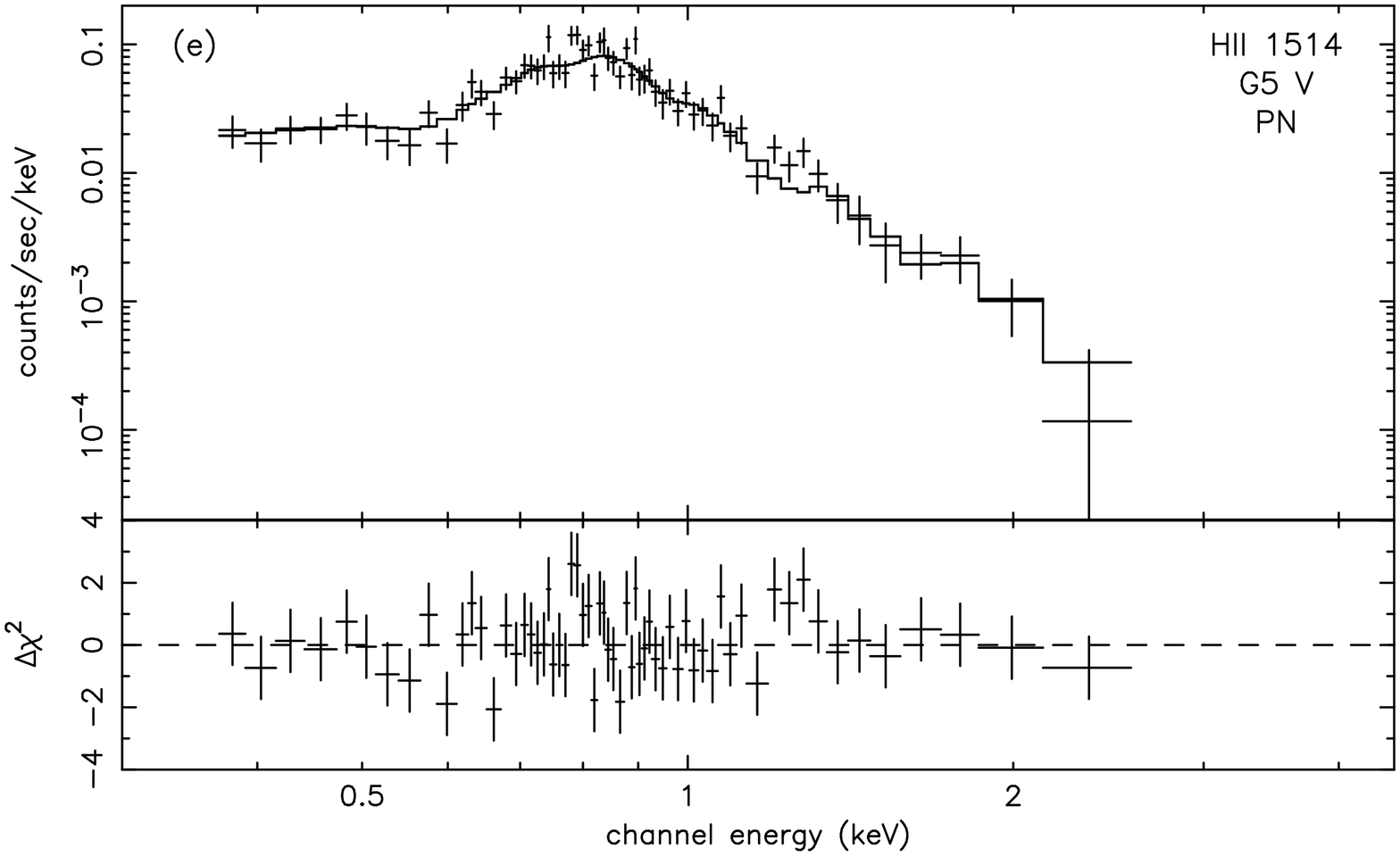}
\end{minipage}%
\begin{minipage}[b]{.45\textwidth}
  \centering
  \includegraphics[height=7.1cm,angle=270]{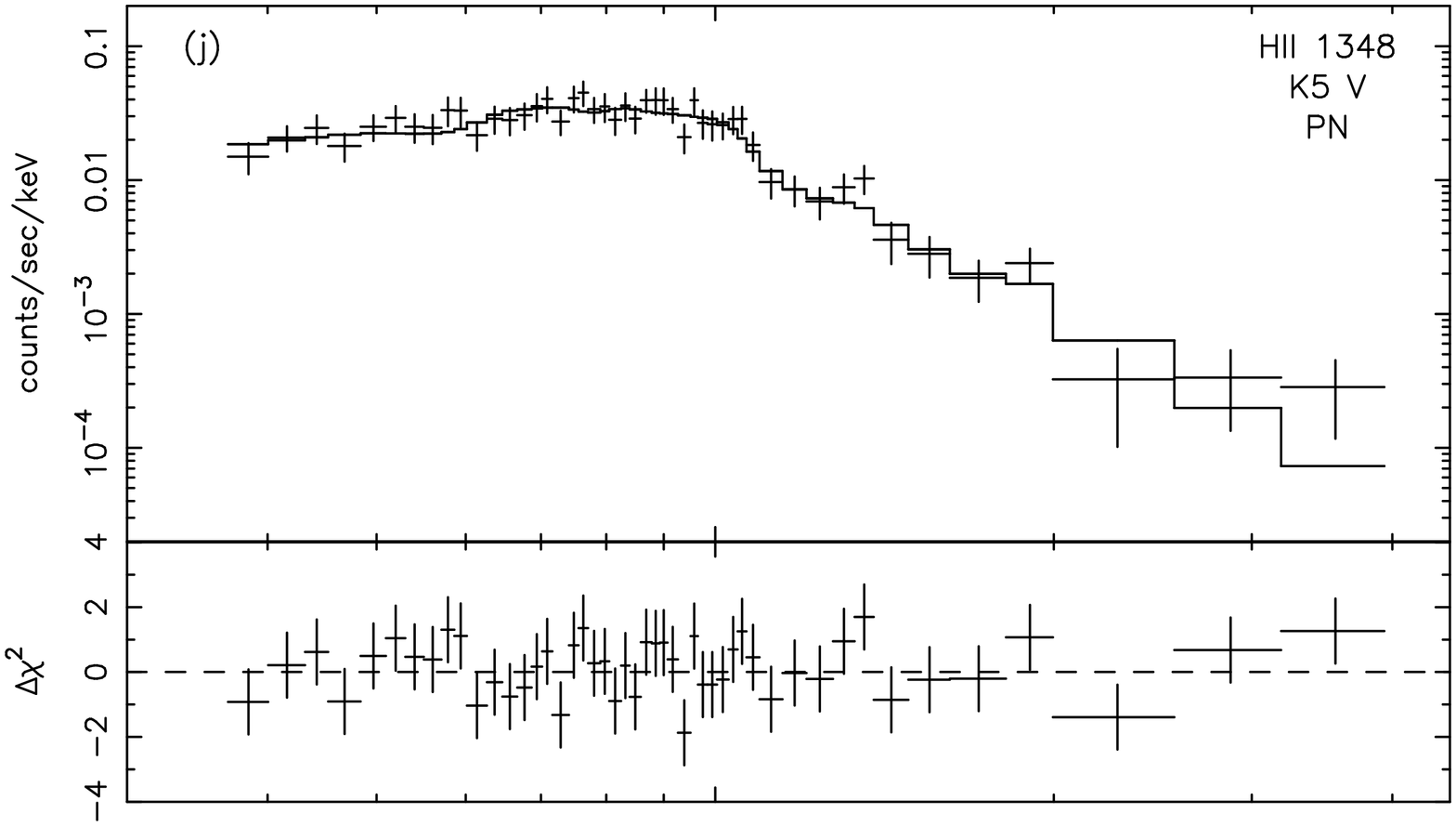}
\end{minipage}\\
\vspace{-0.4cm}
\begin{minipage}[t]{.45\textwidth}
  \centering
\vspace{0.4cm}
\caption{\Quies{} spectra of X-ray-bright Pleiads in the \xmm{} EPIC
field. Each spectrum is labelled with the source name, its spectral
type and the instrument(s) used. In (a), black and grey mark MOS1 and
MOS2 data respectively; in (b) and (f) grey marks the flare peak
spectrum. The displayed model is as detailed in 
Table~\ref{tbl_spec}. The lower panel of each spectrum shows
the \chisq{} residuals. Spectra on the left (a--e) are, or resemble,
those of F or G stars; spectra on the right (f--k) are, or resemble,
those of K stars. Spectra are ordered by \lx, decreasing
from top to bottom.}
\label{fig_spec}
\end{minipage}%
\begin{minipage}[t]{.45\textwidth}
  \centering
  \includegraphics[height=7.1cm,angle=270]{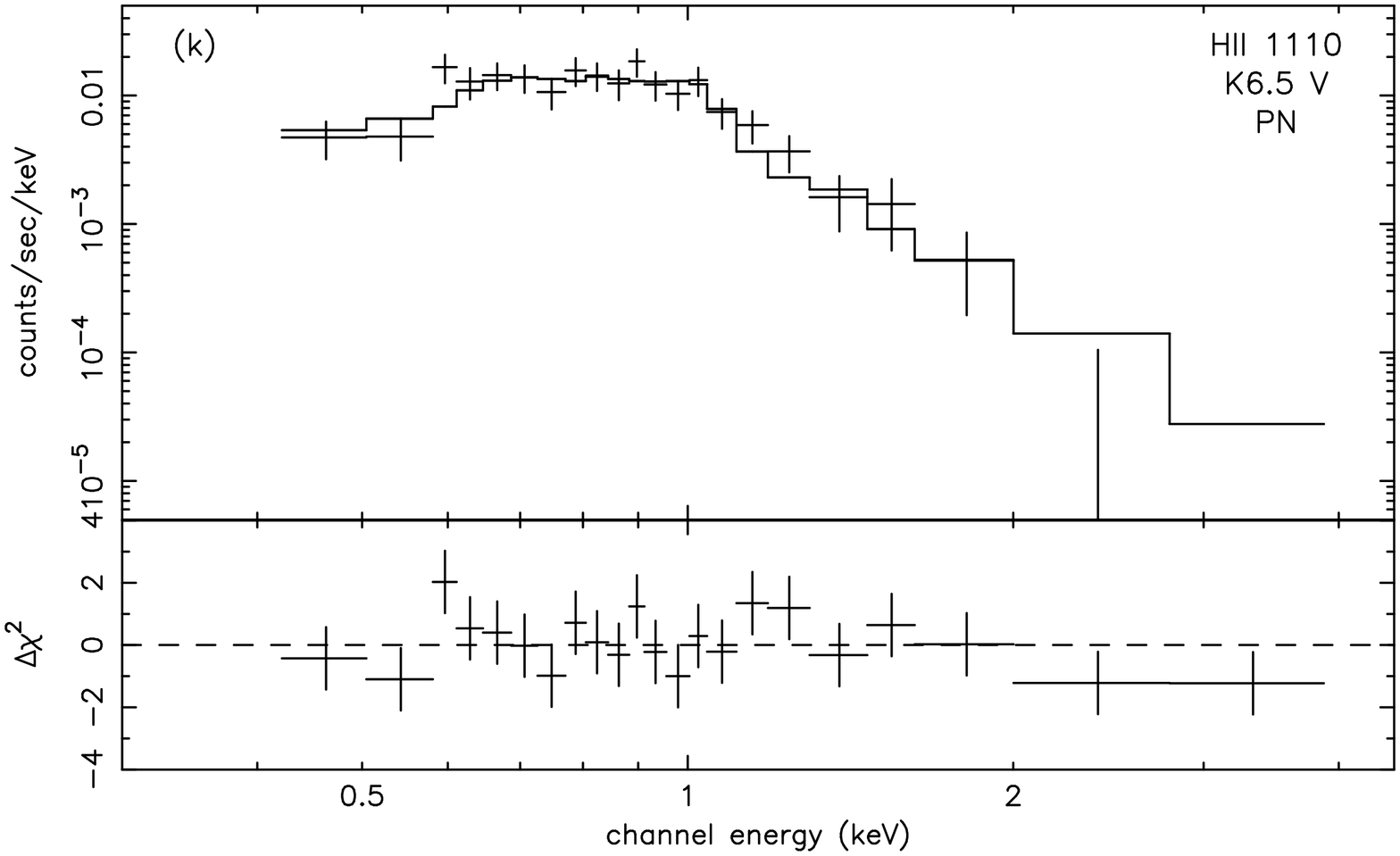}
\end{minipage}\\
}
\end{figure*}

A model was considered to provide an acceptable fit to the data if a
\chisq{} test gave a null hypothesis probability of $\geq 0.10$. The
addition or freeing of extra parameters (i.e. temperature component,
or elemental abundance) was considered to produce a significantly
improved fit if an F-test indicated the probability of an equal or
larger improvement in \chisq{} occurring by chance was $\leq 0.05$. 
The best-fitting models are displayed in Fig.~\ref{fig_spec}, with
parameters, and 90 per cent confidence intervals, listed in
Table~\ref{tbl_spec}.

%%%%%%%%%%%%%%%%%%%%%%%%%%%%%%%%%%%%%%%%%%%%%%%%%%%%%%%%%%%%%%%%%%%%%%%%%%%%%%%
% Figure: Spectra
%%%%%%%%%%%%%%%%%%%%%%%%%%%%%%%%%%%%%%%%%%%%%%%%%%%%%%%%%%%%%%%%%%%%%%%%%%%%%%%

\section{Results and Discussion}
\label{sec_res}

\subsection{\Quies{} emission from solar-like stars}
\label{sec_res_q}

\subsubsection{K-type stars}
\label{sec_res_k}

The five K stars exhibited \quies{} \lx{} and \lxlbol{} values consistent with
those expected (using empirical relations found by Pizzolato \et 2003) from
their rotation periods and spectral types: in particular those of the
binary HII~1100 were consistent with twin K3 components (Bouvier \et
1997), each rotating with period $\sim 7$ d, and the UFRs HII~1516 
and 1280 showed saturated-level emission, although it should be noted
that HII~1348 (Queloz \et 1998) and possibly 1516
(Fig.~\ref{fig_cmd}) are also binary, with uncharacterized companions.

The \quies{} spectra of the five K-stars were remarkably similar:
clearly flatter than those of the earlier-type stars HII~1514 and
1309 (see Fig.~\ref{fig_spec}); requiring 2-T and/or \zfree{}
models for acceptable fits, with $kT_1 \approx
0.35$ and $kT_2 \approx 1.0$ keV (Table~\ref{tbl_spec}), in fair agreement
with the temperatures of 0.28 and 1.10 keV that best-fitted a composite
\ro{} PSPC spectrum of a sample of Pleiades K-stars (GCS95). The two
spectral components have approximately equal emission measures, and
although that of the hotter component is the lower for the two
slowly-rotating stars, HII~1348 and 1110, it is not significantly so.
Fits to the better-quality spectra ($> 900$ net source counts;
HII~1100, 1348 and 1516 -- c.f. $< 500$ counts for HII~1110 and 1280)
were significantly improved by allowing non-solar metallicity
in the 2-T model, with $Z \sim 0.2$ \zsol. DLG02 also
found evidence for low Fe abundances in the spectra of two K-stars in
the \chandra{} sample, and sub-solar coronal metal
abundances appear to be a consistent feature of nearby active 
K-stars (AB~Dor: Mewe \et 1996; ``Speedy'' Mic: Singh \et 1999;
LQ~Hya: Covino \et 2001).

\subsubsection{HII~1032 (G8 V)}
\label{sec_res_fg_1032}

\begin{figure}
\centering{
\includegraphics[height=0.45\textwidth, angle=270]{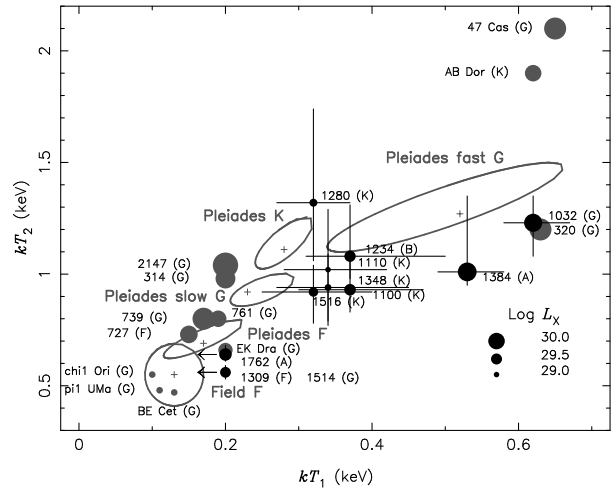}
\caption{
A comparison of coronal temperatures, using 2-T models, of Pleiads and
nearby solar-like field stars. Filled black circles mark Pleiads in the
EPIC FOV (Table~\protect\ref{tbl_spec}); filled grey circles mark
other individual Pleiads and field stars: circle size represents source \lx, and HII number or 
source name, and (spectral type) are given. Open ellipses show
(i) approximate 90 per cent confidence regions for Pleiades samples
divided by spectral type and rotation rate; (ii) approximate limits for a sample of field F-stars. Note the sequence in increasing temperature from F and
slowly-rotating G through K to fast-rotating G stars. The sources
associated with the intermediate-type Pleiads 1762, 1384 and 1234 have
\lx, $kT_1$, $kT_2$ consistent with those of solar-like stars.
}
\label{fig_2t}
}
\end{figure}

HII~1032 exhibited saturated-level X-ray emission, \loglx{} $\approx 30.1$
and \loglxlbol{} $\approx -3.1$, consistent with its rotation period of
1.33 d. The \quies{} spectrum required a 2-T \zfree{} 
model with $kT_1 \approx 0.6$ and $kT_2 \approx 1.3$ keV with a ratio of
hotter:cooler emission measures of $EM_2/EM_1 \sim 1.5$ and metallicity $Z
\sim 0.3$\zsol. A \ro{} PSPC spectrum of HII~1032 was best-fitted
using temperatures of 0.4 and 1.2 keV, with fixed solar metallicity (GCS95).
Comparable best-fitting temperatures were found for a composite PSPC
spectrum of fast-rotating G-type Pleiads, and an 
individual spectrum of the fast-rotating G-star HII~320 (GCS95;
Fig.~\ref{fig_2t}). In mass and saturated-level X-ray 
emission, HII~1032 resembles the nearby, faster-rotating
($P_{\rm rot} = 0.514$ d) K0--1 star AB~Dor,
% also thought to
%be a member of the Pleiades Moving Group (Innis \et 1986), 
but while the \quies{} spectrum of AB~Dor also shows a component 
at 0.6 keV, it has a prominent hot component at 1.9 keV absent on HII~1032 (Mewe \et 1996; Maggio \et 2000; G\"udel \et
2001). The sample of well-studied solar-like UFRs 
is too small for the effect of saturation on coronal properties such
as temperature, to be well-investigated so additions to this sample
are important to our understanding of activity saturation. 
While no companion to HII~1032 has been reported (e.g. Bouvier \et 1997), 
Fig.~\ref{fig_cmd} suggests photometric binarity, so its X-ray
emission may arise from the unsaturated coronae of two or more active stars.

\subsubsection{HII~1514 (G5 V)}
\label{sec_res_fg_1514}

HII~1514 showed \loglx{} $\approx 29.3$, consistent with its rotation
period of 3.33 d. Its spectrum was
acceptably-fitted with a 1-T \zfree{} model with $kT \approx
0.55$ keV and $Z \sim 0.3$\zsol. A second temperature
was not constrained, but a significant component at $kT 
< 0.2$ keV could not be discounted.
Hotter temperatures best-fitted both a composite \ro{} PSPC spectrum
of slowly-rotating G stars ($kT_1 \approx 0.23$, $kT_2 \approx 0.92$
keV), and individual spectra of two reportedly slow-rotating G-stars,
HII~739 and 761 (GCS95), though the \lx{} of these stars ($>10^{30}$
\ergs) indicates \emph{fast rotation} akin to the nearby
solar-analogue EK Dra (G\"udel \et 1997). The slower-rotating
($P_{\rm rot} \approx 5$ d), nearby G1 stars $\pi^1$ UMa and $\chi^1$ 
Ori do show similar \lx{} and temperatures to HII~1514.

\subsubsection{HII~1309 (F6 V)}
\label{sec_res_fg_1309}

HII~1309 gave \loglx{} $\approx 29.4$ and a relatively soft X-ray
spectrum, akin to that of HII~1514, acceptably-fitted with a 1-T,
\zsol{} model with $kT \approx 0.56$ keV, although low metallicity, $Z
\sim 0.5$ \zsol{} offered significant improvement. As for HII~1514, a
significant cool ($kT < 0.2$ keV) component could not be ruled out. 
Similar temperatures ($kT_1<0.2$, $0.4<kT_2<0.7$: mean 0.54 keV) and
\loglx{} (28.5--29.5: mean 29.2) were found for a sample of field
F-stars (Panzera \et 1999), although a composite spectrum of
F-type Pleiads required a hotter temperature of $kT_2 \approx 0.69$
keV (GCS95).
HII~1309's rotation period of $\la 0.74$ d may suggest
saturated-level activity. In late-F stars, saturation appears to set
in at lower \lxlbol{} ($\sim 10^{-4}$) and perhaps shorter rotation
periods than in G-stars (Pizzolato \et 2003).

\subsubsection{Overview}
\label{sec_res_gen}

A rising sequence in coronal temperature ($kT_1$, $kT_2$ in keV) was
observed from the F- and slow-rotating G-stars ($< 0.2$, 0.55), through
the K-stars (0.3, 1.0) to the fast-rotating G-star (0.6, 1.3),
confirming an analysis of composite
\ro{} PSPC spectra of Pleiads (GCS95), which showed the
same trend, albeit with slightly different values, and largely in agreement 
with spectral analyses of field F-stars, using \ro{} (Panzera \et
1999), and nearby solar analogues, using \ro{} and \asca{} (G\"udel
\et 1997; 1998), although none of our 
Pleiades sample has a significant component as hot as 2 keV as do
47~Cas and AB Dor (see Fig.~\ref{fig_2t}).

We also found a requirement or significant preference for
sub-photospheric metallicities in the \quies{} spectra of all solar-like Pleiads
providing $> 500$ source counts: $Z \sim 0.2$\zsol{} for K-stars, $\sim 0.3$\zsol{}
for G-stars and $\sim 0.5$\zsol{} for the F-star. The fitted metallicity,
$Z$, is typically driven by the abundance of Fe, which has strong
emission lines in the EPIC energy range. High-resolution
X-ray spectroscopy of nearby solar-like stars has found coronal
elemental abundances to depend on FIP in a manner that changes with
\lxlbol{} or coronal temperature (Audard \& G\"udel 2002). In
highly-active stars, such as AB~Dor, low-FIP elements such 
as Fe appear underabundant, while in intermediate-activity stars, such as 
$\pi^1$~UMa or $\chi^1$~Ori, there is no FIP bias, and in low-activity 
 stars, such as the Sun, low-FIP elements appear
overabundant (Meyer 1985). The low values of $Z$ we find in the
coronae of the active G- and K-type Pleiads HII~1032, 1100 and 1516
appear to fit into this framework, but the similarly-low $Z$ of the
intermediate-activity HII~1514 does not. Coronal abundances are
entangled with emission measure distribution in spectra of this
quality, and are much better-studied with instruments which can
resolve individual lines and measure \emph{relative} abundances of
different elements, not just a global metallicity parameter.
Overall our results indicate that the temperature structure and
abundances in the \quies{} coronae of solar-like Pleiads conform
to expectations from well-studied nearby active stars. 

\subsection{Variability and flaring on solar-like stars}
\label{sec_res_fl} 

All five K stars exhibited variability at the 90 per cent confidence
level -- only in HII~1348 was the
lightcurve consistent with constant emission with probability $> 0.05$ 
-- and HII~1100, 1516 and 1280 displayed flare-like events
(Figs.~\ref{fig_lc}f, h, i). All three 
flares had decay time-scales, $\tau_{\rm LC}$, of  
$\sim 3$ ks. The HII~1100 flare outshone the rest of its corona five times
over, while the peak \lx{} of the flares on HII~1516 and 1280 
just exceeded those of their respective \quies{} coronae.
The lightcurve of HII~1100 shows a striking dip near the
onset of the large flare that is 
discussed further in \S~\ref{sec_res_fl_dip}.
HII~1032 exhibited a flare (Fig.~\ref{fig_lc}b), with peak \lx{} a
factor $\approx 3$ higher than the \quies{} \lx, and $\tau_{\rm LC} \approx 2$ ks,
while the lightcurves of HII~1309 and 1514 displayed no \emph{formal}
variability (Figs.~\ref{fig_lc}d \& \ref{fig_lc}e). The four flaring stars 
had higher \quies{} \loglxlbol{} values than the four
non-flaring stars (see Table~\ref{tbl_spec}). The results
follow the observed trend for more variability, and more-frequent and
larger flares in the coronae of more-active stars (Stelzer \et 2000).

\subsubsection{Loop modelling from flare decays}
\label{sec_res_fl_loop}

The solar corona is highly structured with hot, X-ray-emitting plasma
largely confined to magnetic loops which may span up to a third of the
solar diameter (e.g. Reale \et 1997).
Although stellar coronae are not spatially resolved by present instrumentation,
it has been shown that the sizes of magnetic loops involved in flares
may be inferred from analysis of the flare decay (van den Oord \& Mewe 
1989; Serio \et 1991). Such techniques have typically neglected the
presence of sustained heating in the decay and yielded loop
half-lengths ${\cal L} > R_{\star}$ in stellar flares (e.g. GCS95). Reale \et
(1997) have performed detailed hydrodynamic modelling of the decay of
flaring loops to develop an empirical method that diagnoses and
accounts for continued heating in the estimation of loop sizes. The
technique has been successfully tested on resolved solar flares and
applied to stellar flares using a number of different X-ray detectors
(including EPIC \pn), yielding ${\cal L} < R_{\star}$ (e.g. Reale \& Micela
1998; Favata \& Schmitt 1999; Maggio \et 2000; G\"udel \et 2001). The
presence of sustained heating is diagnosed through the evolution of
temperature, $T$, and density, $n$, in the flare decay, specifically
the shallower the slope, $\zeta$, in the $\log T$ -- $\log n$
diagram, the slower the decay time-scale of the flare heating rate
(maximum $\zeta \approx 1.9$ for no continued heating; EPIC \pn). For
stellar flares $n$ can rarely be directly measured so the emission
measure, $EM = \int n_e\,n_H\,dV \propto n^2$, is used to obtain
$\zeta$ from a $\log T$ -- $\log \sqrt{EM}$ diagram.
${\cal L}$ is then inferred from $\zeta$, the lightcurve decay time,
$\tau_{\rm LC}$, and the maximum temperature in the
loop at flare peak, $T_{\rm max}$, as in equations (4) and (5) in
Favata \et (2000), where $c_A = 11.6$, $\zeta_A = 0.56$, $q_A = 1.2$
for EPIC \pn{} (Reale, priv. comm.).
%: 
%\[
%{\cal L} = {{\tau_{\rm LC} \sqrt{T_{\rm max}}}/{3.7 \times 10^{-4} F(\zeta)}}{\rm .}
%\]
%\[
%{\rm where\ } F(\zeta) \approx  11.6 e^{-\zeta/0.56} + 1.2 {\rm \ (for\ EPIC\ \pn)}
%\]

\begin{figure}
\centering{
\includegraphics[height=0.45\textwidth, angle=270]{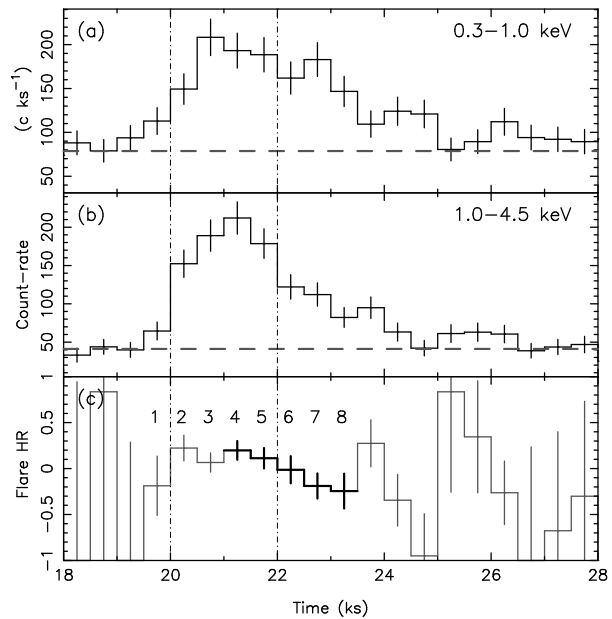}
\caption{
Time-series for the flare exhibited by HII~1032: (a) and (b) 
background-subtracted count-rates in the 0.3--1.0 and 1.0--4.5 keV
bands respectively; (c) hardness ratio, defined in the 
text. The mean \quies{} value of count-rate in each band 
%(calculated from times outside the plotted time-interval) 
is shown by dashed
grey lines. The points used for analysis of the flare decay are 
shown in bold; numbers refer to points in Fig.~\ref{fig_zeta}. The
interval used for the extraction of the flare-peak spectrum is bounded
by dot-dashed lines.
}
\label{fig_hr}
}
\end{figure}

The flares on HII~1032 and 1100 allow us to apply a modified version
of the hydrodynamic technique to estimate the size of loop involved.
For each flare, a spectrum of the flare ``peak''
(displayed in grey in Figs.~\ref{fig_spec}b and \ref{fig_spec}f)
was extracted and fitted with a single \mekal{}
component with $Z$ fixed to the best-fitting \quies{} value, in
addition to the best-fitting \quies{} model
(from Table~\ref{tbl_spec}). The fitted temperatures (with \onesig{}
confidence intervals) and emission
measures of the plasma at flare peak were 23.5 (20--27)
MK and $3.4 \times 10^{53}$ cm$^{-3}$ for HII~1032, and 15.4 (14--17)
MK and $1.4 \times 10^{53}$ cm$^{-3}$ for HII~1100, from which we
derived respective $T_{\rm max}$ values of 39.2 (32--46) and 
24.3 (22--27) MK using the relation (Reale, priv. comm.):
\[
T_{\rm max} = 0.184\:T_{\rm obs}^{1.13} {\rm \ (for\ EPIC\ \pn).}
\]

\begin{figure}
\centering{
\includegraphics[height=0.4\textwidth, angle=270]{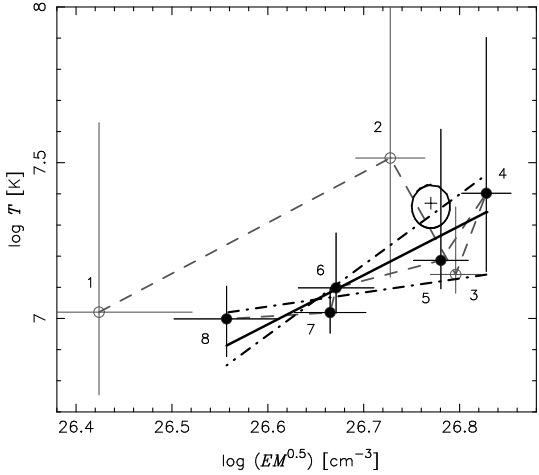}
\caption{
The evolution of estimated temperature ($T$) and electron
density ($\propto \sqrt{EM}$) during the HII~1032 flare. $T$ and $EM$
were estimated from hardness ratio and count-rate. Dashed grey lines
connect sequential points; numbers refer to bins in Fig.~\protect\ref{fig_hr}. The slope, $\zeta$, is a diagnostic for sustained heating
through the flare decay. The points used to estimate $\zeta$ are shown
in bold. The bold slope shows the best fit; the dot-dashed slopes are
approximate \onesig{} limits. The ellipse shows the approximate
\onesig{} confidence interval for parameters derived from the
flare-peak spectrum.
}
\label{fig_zeta}
}
\end{figure}

We constructed a $\log T$ -- $\log \sqrt{EM}$ diagram
by estimating the flare temperature from a hardness ratio ($HR =
(B-A)/(A+B)$ where $A$ and $B$ are the count-rates above the 
mean \quies{} count-rate in the 0.3--1.0 and 1.0--4.5 keV bands
respectively), and then the emission measure for that temperature from
the total count-rate $A+B$. Conversions were calculated using the \pn{}
 response appropriate for the particular star
(\S~\ref{sec_xdata_spex}) and 1-T \mekal{} emission models with
$Z$ fixed to the best-fitting \quies{} value. The count-rates were
not sufficient to constrain $\zeta$ well: $\zeta_{1032}$ = 1.6
(0.45--2.3); $\zeta_{1100}$ = 1.0 ($< 1.6$). Although in the HII~1032
flare $\zeta$
is consistent with the absence of sustained heating, its presence
cannot be ruled out, and it is apparently required in the case of
HII~1100. If $\zeta < 0.4$, the observed decay is controlled by
the heating decay time-scale, not the thermodynamic time-scale, and is
hence independent of ${\cal L}$, which can be constrained only by an
upper limit: this is the case for HII~1100. To estimate the lower and
upper bounds of a \onesig{}
confidence interval for ${\cal L}$ we have first calculated the lower and upper bounds for the
quantity ${\cal L}\,F(\zeta)$ using the \onesig{} uncertainties on $T_{\rm
max}$ and $\tau_{\rm LC}$, secondly added in quadrature a 20 per cent
uncertainty that derives from consistency checks on solar flares, and
finally used the lower and upper bounds 
of $\zeta$ to obtain upper and lower bounds respectively to 
$F(\zeta)$. We thus estimate ${\cal L}_{1032} \sim 1.8$ (0.4--2.6) $\times 10^{10}$ cm $\sim
0.3$ (0.07--0.45) $R_{\star}$ and ${\cal L}_{1100} \sim 1.2$ ($< 2.5$) $\times 10^{10}$ cm $\sim
0.25$ ($< 0.5$) $R_{\star}$. Assuming a loop aspect ratio of
0.2, the mean electron densities in
the loop at flare peak, $n$, were 
$n_{1032} \sim 5$ (2.7--45) and
$n_{1100} \sim 5$ ($> 2$) $\times 10^{11}$ cm$^{-3}$. The
corresponding pressures, $p$, calculated from the equation of state, were
$p_{1032} \sim 5$ (2.5--60) and
$p_{1100} \sim 4$ ($> 1$) $\times 10^3$ dyne cm$^{-2}$, requiring
magnetic field strengths, $B$, of $B_{1032} \sim 400$ (250--1200)
and $B_{1100} \sim 300$ ($> 170$) G for confinement.
These values are fairly typical of a variety of larger flares on the nearby young
K-type stars AB~Dor (${\cal L}$ in the range 2.5--$4.9 \times 10^{10}$ cm:
Ortolani \et 1998; Maggio \et 2000; G\"udel \et 2001), and LQ~Hya
(${\cal L} \approx 8.3 \times 10^{10}$ cm: Covino \et 2001) that have been
analysed using the hydrodynamic method. 
%The expected pressures for a loop in equilibrium, calculated from the
%RTV scaling laws are lower than the pressures calculated above, but
%there are large uncertainties in the values; one would expect $p_e$
%not to reach equilibrium values.

We caution that {\it TRACE} images of large solar flares 
reveal structure far more complex than single loops (e.g. review by
Aschwanden, Poland \& Rabin 2001), and so stellar flares, often $10^{2 - 3}$
times more powerful than the strongest solar flares, may not be
well-modelled as single loops.

\subsubsection{An eclipse of the flare on HII~1100?}
\label{sec_res_fl_dip}
The lightcurve of HII~1100 displays a striking dip at 24--25.5 ks
after the start of the pn exposure when the count-rate drops sharply
from $\approx 
120$ c ks$^{-1}$ to the \quies{} rate of $\approx 40$ c ks$^{-1}$ in
$\tau_{\rm fall} \approx 300$ s, remains at this value for $\tau_{\rm min} \approx 600$ s, and then
rises sharply to $\approx 200$ c ks$^{-1}$ in $\tau_{\rm rise} \approx 500$ s
(Fig.~\ref{fig_lc}f). The feature is not of instrumental
origin as it is reproduced in the MOS lightcurves, the
background on that CCD shows no similar feature and no periods of 
data-loss at this time were recorded.

The profile of the dip is suggestive of a total eclipse of a slow-rising
flare, and the most interesting possibility is that this is caused by
an object orbiting the flaring star. Approximating the flare 
geometry to the simplest case of a uniform circular source, and
assuming a near-central eclipse, the ratio of radii of eclipsing
object, $R_{\rm b}$, and flare, $R_{\rm fl}$, is:
\[
R_{\rm b}/R_{\rm fl} = 1 + \tau_{\rm min}/\tau_{\rm fall} \approx 3
\]
\[
{\rm while\ } R_{\rm b} = (\tau_{\rm fall} + \tau_{\rm min}) v / 2
\]
\noindent where $v$ is the orbital velocity of the eclipsing
object. Substituting for $v$ using Kepler's 3$^{\rm rd}$ Law, the
product of the circular orbital distance, $a_{\rm b}$, and $R_{\rm b}^2$ is:
\[
a_{\rm b} R_{\rm b}^{2} = (\tau_{\rm fall}+\tau_{\rm min})^{2} G
M_{\star}(1+M_{\rm b}/M_{\star})/4 \approx 22
\]
\noindent where $a_{\rm b}$ and $R_{\rm b}$ are measured in $10^{10}$
cm and $M_{\star} = 0.8 M_{\odot} \gg M_{\rm b}$.
The loop half-length, ${\cal L} = \pi R_{\rm fl} / 2 \sim 1.2 \times 10^{10}$
cm, inferred from the flare decay suggests $R_{\rm b} \sim 2.3 \times 10^{10}$
cm, and, unacceptably,  $a_{\rm b} \sim 4 \times 10^{10}$
cm, $< R_{\star}$, although the geometry of the loop and its
inclination may cause $R_{\rm b}/R_{\rm fl} < 3$, and a smaller loop
size was allowed by the flare-decay analysis. A Jupiter-sized object
($R_{\rm b} = R_{\rm J} \approx 0.7 \times 10^{10}$ cm) must
orbit at a distance of $\sim 0.03$ AU to cause the eclipse, and
would imply a flare-loop half-length of ${\cal L} \sim 0.4 \times 10^{10}$ cm, 
the flare decay hence being controlled by the heating decay time-scale, 
independent of the loop size. A number of such ``hot Jupiters'' have
been detected around nearby solar-like stars (e.g. 51 Peg b: Mayor \& Queloz
1995; $\tau$ Bo\"o b: Butler \et 1997). Although HII~1100 is a binary
system, the projected separation of the near-twin components of 100 AU
(Bouvier \et 1997) is too large to affect the orbit of such a planet. 
The existence 
of such a planet could be demonstrated by optical observations. Its
transit across the stellar disc should eclipse a fraction $(0.7/5.0)^2
= 0.02$ of the disc area, causing a dimming of 0.02 mag. Given the
above orbit, this dimming should last $\sim 3$ ks and occur every
$\sim 2$ d. The gravitational influence of the planet should also
cause a small periodic radial velocity perturbation of $0.2\, M_{\rm
b} \sin i /M_{\rm J}$ \kms, although HII~1100's apparent magnitude of
$V = 12.16$ makes such measurements extremely difficult. If a planet
were to be confirmed, this would be the first planet detected by its eclipse of
an X-ray flare, the first planet discovered in the Pleiades, and only
the second planet discovered by a transit method. However, one must
conclude that even if such a planet was to exist, its
total eclipse of a flaring loop would be a highly fortuitous event.

An alternative possibility is that the flare was obscured by cool
material passing across the line of sight, as proposed by Haisch \et
(1983) to explain the observation by \ei{} of a temporarily increased \nh{}
for a time interval during a flare on the dMe star Proxima
Centauri. If we model the flare plasma using the best-fitting spectral
model of the flare peak (\S~\ref{sec_res_fl_loop}), an absorbing
column of \nh{} $> 10^{22}$ cm$^{2}$ is required to absorb $> 75$ per
cent 
of the 0.3--4.5 keV photons, as is observed. This is two orders 
of magnitude greater than has been seen in solar prominences, the
Proxima Centauri event, or cool clouds in the atmosphere of AB Dor
that have been inferred from transient H$\alpha$ absorption profiles
(Collier Cameron \et 1990). While the
size of the AB Dor clouds (area $\sim 10^{21}$ cm$^{2}$) is comparable
to that required by our eclipse profile, their velocities, held to
the rotation of the star at around the corotation radius, $\sim 200$--300
\kms, are not analogous to the 
velocity of $\sim$ 150--500 \kms{} required of our obscuring cloud,
 as \vsini{} is only 5 \kms{} for HII~1100. While such
high velocities are fairly common in erupting prominences on the Sun
(e.g. Gopalswamy \et 2003), and the bulk kinetic energy of the cloud
is around the same order of magnitude as (0.05--5 times) the total
energy, $\sim 6 \times 10^{33}$ erg, radiated by the flare with which
it is presumably associated, it would be puzzling if such a flare,
which is a routine twice-a-day occurance on AB Dor, was associated
with such an extraordinarily denser cloud.

A third explanation is that the large flare, with an impulsive, short
rise-time, is shortly preceeded by a smaller flare (albeit, with peak
\lx{} $\approx 10^{30}$ 
\ergs, still larger than the flares exhibited by HII~1516 and
1280) with a shorter decay time than rise time. Evidence for a further flare-like event occurring 
during the flare decay, at $\approx 27.8$ ks (see Fig.~\ref{fig_lc}f)
supports the idea of multiple flaring within a complex active
region. If we apply the hydrodynamic model with $T_{\rm max} \sim 25$ MK,
the fast decay, $\tau_{\rm fall} \sim 300$ s, 
indicates a loop structure, ${\cal L} \la 10^{9}$ cm $\la 0.02 R_{\star}$,
smaller than 
that of the larger flare that followed, though similar to that
inferred from a flare with peak \lx{} $\approx 4 \times 10^{29}$
\ergs, observed on a component of the dMe binary YY Gem (Stelzer \et 2002). 
Flares with a longer rise than decay time do not form a
well-recognized class on the Sun, but they have been occasionally
observed on other stars (e.g. the dMe star EQ Peg: Pallavicini,
Tagliaferri \& Stella 1990; several examples in the Pleiades and
Hyades noted by Stelzer \et 2000). 
The coincidence of two flares -- the first unusual, the second
impulsive -- in such a way as to mimic an eclipse forms a somewhat
{\it ad hoc} explanation of the feature, but none of the available
hypotheses presents a convincing explanation on the grounds of
probability and/or precedent. A planetary eclipse is the most
interesting possibility and the only one that may be confirmed or
refuted by further observations.

\begin{table*}
\begin{minipage}{\textwidth}
\caption{Physical and X-ray characteristics of intermediate-type Pleiads in
the \chandra{} and \xmm{} fields. Columns show: 
(2) spectral type of primary; 
(3) \vsini{} in \kms; 
(4) and (5) binarity flag and reference: 
%VB = visual binary;
%SB1 = single-lined spectroscopic binary (SB); 
%SB2: double-lined SB;
%SB2O: SB2 with calculated orbit; 
%SB?: suspected SB; 
%(5) reference for binarity: 
1. Abt \et 1965;
2. Liu \et 1991; 
3. Mason \et 1993; 
4. H{\o}g \et 2000; 
5. Dommanget \& Nys 2000; 
(6) and (7) estimated primary and secondary masses (Raboud \&
Mermilliod 1998); 
(8) spectral type of secondary, inferred from $M_2$;  
(9) field; 
(10) and (11) best-fitting temperatures in keV for X-ray spectrum; 
(12) \loglx{} in the 0.5--2.0 keV band; 
(13) proposed spectral type of dominant X-ray source. 
\chandra{} data is from DLG02.
}
%\vspace{0.1cm}
\label{tbl_early}
\small
\centering{
\begin{tabular}{rlrlrcccrrrrc}
\hline
HII & 
\multicolumn{1}{c}{SpT$_{\rm A}$} & 
\multicolumn{1}{c}{\vsini} &
\multicolumn{1}{c}{Bin.} & 
\multicolumn{1}{c}{Ref.} & 
$M_{\rm A}$ & 
$M_{\rm B}$ &
SpT$_{\rm B}$ &
\multicolumn{1}{c}{X-ray Obs} & 
\multicolumn{1}{c}{$kT_1$} & 
\multicolumn{1}{c}{$kT_2$} & 
\multicolumn{1}{c}{\loglx} & 
SpT(X)\\
\multicolumn{1}{c}{(1)} & \multicolumn{1}{c}{(2)} &
\multicolumn{1}{c}{(3)} & \multicolumn{1}{c}{(4)} &
\multicolumn{1}{c}{(5)} & \multicolumn{1}{c}{(6)} &
\multicolumn{1}{c}{(7)} & \multicolumn{1}{c}{(8)} &
\multicolumn{1}{c}{(9)} & \multicolumn{1}{c}{(10)} &
\multicolumn{1}{c}{(11)} & \multicolumn{1}{c}{(12)} &
\multicolumn{1}{c}{(13)}\\
\hline
1309 & F6 V & 85 & & & 1.21 & & & \xmm & $<0.2$ & 0.56 & 29.4 & F\\
1122 & F4 V & 28 & SB2 & 2 & 1.22 & 0.51 & K/M & \chandra & -- & 0.45 & 29.1 & 
F\\
1338 & F3 V & 10 & SB2 & 2 & 1.22 & 1.22 & F & \chandra & -- & -- & 28.7 & F\\
1762 & A9 V & 180 & SB2, VB & 2, 3 & 1.36 & 0.86 & G/K & \xmm & $<0.2$ & 0.64 &
29.6 & G\\
1284 & A9 V & 100 & SB1 & 2 & 1.45 & 0.86 & G/K & \chandra & -- & -- & 27.6 & K\\
956 & A7 V & 150 & SB?, VB & 2, 4 & 1.37 & 1.18 & F & \chandra & -- & 0.57 &
29.3 & F\\
1362 & A7 V & $< 12$ & & & 1.52 & & & \chandra & -- & -- & $< 27.9$\\
1384 & A4 V & 215 & & & 1.61 & & & \xmm & 0.53 & 1.01 & 30.1 & G\\
1028 & A2 V & 110 & VB & 3 & 1.86 & 1.45 & A & \xmm & -- & -- & $<27.6$\\
1431 & A0 V & 40 & SB2 & 1 & 1.97 & 0.68 & K & \xmm & -- & -- & $<27.6$\\
1375 & A0 V & 160 & SB1 & 2 & 2.11 & 1.74 & A & \chandra & -- & -- &
$<28.0$\\
1234 & B9.5 V & 260 & VB & 3 & 2.11 & 1.33 & A/F & \xmm & 0.37 & 1.08 &
29.4 & K\\
980 & B6 IV & 275 & VB & 5 & 2.53 & ? & ? & \chandra & -- & 0.56 & 29.6 & G\\
\hline
\end{tabular}
}
\end{minipage}
\end{table*}

\subsection{Emission from intermediate-type stars?}
\label{sec_res_early}

Intermediate-type stars ($\approx$B4--F4) are believed
to lack both the massive stellar wind and deep convective envelope
intrinsic to the X-ray generation mechanism of early-type and
solar-like stars respectively. Indeed, most X-ray surveys of A-type
stars in the field and in open clusters have found a paucity of
detections (e.g. Schmitt \et 1985; Micela \et 1990; Schmitt \et
1990). The \lx{} of detected stars appears to be uncorrelated
with stellar properties influential in the emission mechanisms of
early-type or solar-like stars, such as \lbol, or rotation
(Simon, Drake \& Kim 1995; Panzera \et 1999). However, chromospheric
emission has been reported to show a correlation with rotation rate
for stars with spectral types as early as A9 (Schrijver 1993), and
emission lines of C III and O VI in $\tau^3$ Eri indicate material at
temperatures in excess of 0.3 MK is possible in the atmosperes of stars as early as A4
(Simon \et 2002). In the
Hyades, a sharp fall in chromospheric emission is observed to occur at
spectral type F5, which coincides with a sharp fall in rotation
velocity (B\"ohm-Vitense \et 2002). It is possible that dynamo-driven
coronal X-ray emisson extends to spectral types some subclasses
earlier than F5. 95 percent of all F-type stars within 13 pc of the
Sun were detected by \ro{} (Schmitt 1997), although only an F2
subgiant and an unresolved pair of F0 V stars were earlier than F5. $\alpha$
Aqu (Altair) and $\alpha$ Cep were the only assuredly-single A-stars
in the solar neighbourhood to be detected; both are
 fast-rotating, A7 stars with very cool, low-luminosity, coronae
(\vsini{} $> 210$ \kms, $kT \approx 0.1$ keV, \lx{} $\approx 2 \times
10^{27}$ \ergs: Simon \et 1995). However, detected intermediate-type stars
have usually had hotter temperatures and \lx{} of
factors up to $10^{2-3}$ higher; similar to the most active solar-like
stars. Hence, such detections have
been conventionally attributed to coronal emission from late-type
companions (e.g. Golub \et 1983; Caillault \& Zoonematkermani 1989;
Caillault, Gagn\'e \& Stauffer 1994; Micela \et 1996). In some cases
further investigation has uncovered an active late-type companion
(G\"udel \et 1998; Simon 
\& Ayres 2000) but often no further evidence for an appropriate
partner exists, and in \ro{} HRI studies of resolved late-B + late-type
(\emph{Lindroos}) binary systems, the B-type component was found
to be the brighter X-ray source in many cases (Bergh\"ofer \& Schmitt
1994). However, one cannot rule out the existence 
of further late-type companions closer still to the primary, and
hardness ratios of these X-ray-bright B-type stars are consistent with
those of late-type coronae (Hu{\' e}lamo \et 2000). 

Three of the five intermediate-type stars in our observation were detected. 
We consider available optical and binarity information for these
stars, in addition to our own X-ray spectral and timing data 
(Table~\ref{tbl_early}), to test for consistency with a late-type companion.
Where a companion is known, we have inferred an approximate spectral type from
the estimated masses of Raboud \& Mermilliod (1998).

\subsubsection{HII~1762 (A9 V)}

HII~1762 was close to the edge of the EPIC FOV 
and so only observed by MOS1. Its lightcurve showed no significant
variability. Its X-ray spectrum was best-fitted by a
1-T \zsol{} model with $kT \approx 0.64$ keV and \loglx{} $\approx
29.6$: both somewhat higher than found for HII~1514 and 1309. HII~1762 has a
known spectroscopic and visually-separated companion of estimated
spectral type G--K: a G-star more active than HII~1514 is a viable
source of the X-ray-emission.  

\subsubsection{HII~1384 (A4 V)}

HII~1384 is one of 
the brightest X-ray sources in the Pleiades with \loglx{} $\approx 30.1$
reported in \ei{}, \ro{} and \xmm{} observations. A flare with
peak \loglx{} of 30.5 and decay time of $\approx 1.2$ ks was noted by
Stelzer \et (2000).

HII~1384 fell on the very edge of a \pn{} CCD, which may have caused the
variability not observed in the MOS lightcurves. Analyses of spectra
from the \pn, and from MOS1, and simultaneous fitting of the two
MOS spectra were in good agreement, but the MOS2
spectrum required a hotter second component. A \zfree{} model was
required to give an acceptable fit and the addition of a second
component gave significant improvement. The spectral parametrization: 
$kT_1 \approx 0.5$, $kT_2 \approx 1.0$ keV, with $EM_2/EM_1 \sim 0.8$,
and $Z \sim 0.25$ 
\zsol, is similar to that of coronae on fast-rotating G-stars
(Fig.~\ref{fig_2t}), although no companion to HII~1384 has been found. Radial 
velocity variations of $\sim 35$ \kms{} reported by Abt \et (1965)
have been doubted (Simon \et 1995; Raboud \& Mermilliod 1998), and
more recent studies, albeit based on just two measurements each, found 
no evidence for variations indicative of a close companion (Liu
\et 1991; Morse, Mathieu \& Levine 1991). Speckle interferometry
revealed no companion at angular separations of 0.035--1 arcsec unless 
it is $> 3$ mag. fainter than the $V=7.66$ primary (Mason \et 1993),
so a late-G star could have escaped detection. 
Fig.~\ref{fig_cmd} does suggest photometric binarity. The
apparently-extraordinary X-ray emission of HII~1384 could be explained 
by a highly-active G-type companion, just like those discovered to
accompany 47~Cas (G\"udel \et 1998) and the Hyad 71~Tau (Simon \&
Ayres 2000), two F0 stars with apparent \lx{} of $2 \times 10^{30}$
\ergs. A dedicated radial-velocity study of the star could test this
hypothesis.

\subsubsection{HII~1028 (A2 V)}
HII~1028 showed no X-ray emission above a threshold \loglx{} of
27.6. A visual companion, estimated to be a late-A star,
is also not expected to be a strong X-ray emitter.

\subsubsection{HII~1431 (A0 V)}
HII~1431 also showed no X-ray emission above a threshold \loglx{} of
27.6. This is somewhat surprising as it has a spectroscopic companion 
estimated to be a K-star whose X-ray emission would have to be among
the faintest 5 per cent for K-stars in the Pleiades according to the
luminosity functions of Stauffer \et (1994) and Stelzer \& Neuh\"auser
(2001).

\subsubsection{HII~1234 (B9.5 V)}
HII~1234 showed a strong increase in X-ray output of factor $\sim 2$
toward the end of the observation (Fig.~\ref{fig_lc}g). The
\quies{} spectrum required a second component or \zfree{} for an
acceptable fit, and significant improvement was found using a 2-T
\zfree{} model. The \loglx{} of 29.5 and spectral parametrization:
$kT_1 \approx 0.4$, $kT_2 \approx 1.1$ keV, with $EM_2/EM_1 \sim 1.5$
and $Z \sim 0.25$ \zsol, 
resemble those for an active K-star (Fig.~\ref{fig_2t}). 
Spectral parametrization of the higher
emission state was poorly-constrained but indicated higher
temperatures ($\sim 0.7$ and 2 keV), similar abundances and
\loglx{} $\approx 29.7$. HII~1234 has a visual companion also of
intermediate-type and not expected to be a strong X-ray source. The
X-ray emission would be better-explained by a hidden K-type companion
in the system.

\subsubsection{The combined \chandra{} and \xmm{} sample}
X-ray emission from the detected intermediate-type stars
in the \xmm{} observation appears consistent with coronal emission
from companions of spectral types F, G or K. The inclusion of
intermediate-type Pleiads in the \chandra{} field analysed by DLG02
 more than doubles the sample size
(Table~\ref{tbl_early}). Our interpretation of the results from the
combined sample is that 
the soft X-ray emission from the three F-type stars is probably dominated
by coronal emission from the primaries (or the F-type companion in
HII~1338), while the strong, soft X-ray emission of the A and B stars HII~956,
980 and 1762 is not intrinsic (as proposed by Krishnamurthi \et 2001),
but due to F-, or moderately-active G-type companions, which are
probably identical with companions already noted in the 
literature. Conversely, the companions of HII~1431 and
1284 are unusually weak X-ray-emitters if they are truly
K-type stars. The strong, harder emissions of HII~1234
and 1384 are plausibly due to highly-active K- and G-type
companions respectively, but neither have yet been 
discovered. Dedicated radial velocity and interferometric studies
are required to confirm the presence of such companions, in tandem with a
wider \xmm{} and/or \chandra{} X-ray spectroscopic survey of
intermediate-type Pleiads to increase the sample size. 
If it can be firmly demonstrated that X-ray emission from
intermediate-type stars is solely due to companions, X-ray
studies may become an effective method of not only surveying the
binarity status of stars in this mass range, but also constraining 
the spectral types of companions. 

\section{Summary}
We have performed detailed spectral and timing
analyses of a sample of 13 intermediate-type and solar-like Pleiads,
contained in a 40-ks \xmm{} EPIC observation of the core of the
cluster, that enabled the study of individual members with X-ray luminosities
an order of magnitude lower than achieved using \ro. 
All solar-like members (1 F-, 2 G- and 5 K-stars) had \lx $\ga
10^{29}$ \ergs{} and consistent
with a rotation--activity connection: three ultra-fast rotators (UFRs)
emitted at a saturated level. Variable
emission was observed from the K-stars and fast-rotating
G-star, and the four with highest \quies{} \lxlbol{} values exhibited
flares. Hydrodynamic modelling of the flares on HII~1032 and 1100
led us to infer loop structures of half
lengths ${\cal L} < 0.5\, R_{\star}$. The lightcurve of HII~1100
showed a dip feature that could be due to an eclipse by a ``hot
Jupiter''-like planet, absorption by a cool prominence with \nh $>
10^{22}$ cm$^{-2}$ moving across the line of sight at several hundred
\kms, or the coincidence of two flares, the first with an unusual
decay, faster than its rise, indicating a small loop, ${\cal L} <
0.02\, R_{\star}$. None of these explanations is satisfactory in
likelihood or precedent, but the existence of a planet may be tested
by optical photometric monitoring of the star for transits with depth of
$\approx 0.02$ mag, followed by a spectroscopic search for radial velocity
variations, though the star is faint ($V = 12.16$). 
Spectral modelling of the
\quies{} X-ray emission using two-temperature models: (a) 
confirmed the temperature sequence with spectral type (in order of rising
temperature: F, slowly-rotating G, K, fast-rotating G) deduced from
composite spectra of Pleiades samples using the lower-resolution \ro{}
PSPC, and (b) indicated sub-solar metallicities. The results
show that solar-like Pleiads conform to expectations from
well-studied active solar-like stars, although none of the
three UFRs had the hot, 2 keV component seen in the
saturated coronae of the nearby stars AB~Dor and 47~Cas.
Three of the five intermediate-type Pleiads showed strong X-ray
emission consistent with coronal emission from a solar-like
companion. For such systems, the X-ray spectrum gives useful insight
into the spectral type of the companion. 

\section*{Acknowledgments}
KRB and JPP acknowledge the financial support of the UK Particle
Physics and Astronomy Research Council (PPARC). 
The authors would like to thank F.~Reale for guidance in applying
hydrodynamic flare decay modelling, M.~G\"udel for helpful discussions, 
J.~Linsky for suggesting absorption as a possible origin of the
`eclipse', A.~Collier Cameron, M.~Audard and B.~Stelzer for useful
input, and the anonymous referee whose comments helped to greatly
improve the clarity of the paper. 
This work uses data obtained by \xmm{}, an ESA science
mission with instruments and contributions directly funded by ESA
Member States and the USA (NASA), and
made use of archival material from the SIMBAD and VIZIER systems at
CDS, Strasbourg, NASA's Astrophysics Data System, and the Leicester
Database and Archive Service (LEDAS).
%
%The Digitized Sky
%Survey was produced at the Space Telescope Science Institute 
%under U.S. Government grant NAG W-2166. The images of
%these surveys are based on photographic data obtained using the Oschin Schmidt
%Telescope on Palomar Mountain and the UK Schmidt Telescope. The plates were
%processed into the present compressed digital form with the permission of these
%institutions.
%

{}

\label{lastpage}

\end{document}